# Infrared Spectra, Optical Constants and Temperature Dependences of Amorphous and Crystalline Benzene Ices Relevant to Titan

Short title

Benzene ice infrared spectra relevant to Titan

Delphine Nna-Mvondo[1][2][3], Carrie M. Anderson[3]

## ABSTRACT

Benzene ice contributes to an emission feature detected by the Cassini Composite InfraRed Spectrometer (CIRS) near 682 cm$^{-1}$ in Titan's late southern fall polar stratosphere. It is as well one of the dominant components of the CIRS-observed High Altitude South Polar (HASP) ice cloud observed in Titan's mid stratosphere during late southern fall. Titan's stratosphere exhibits significant seasonal changes with temperatures that spatially vary with seasons. A quantitative analysis of the chemical composition of infrared emission spectra of Titan's stratospheric ice clouds relies on consistent and detailed laboratory transmittance spectra obtained at numerous temperatures. However, there is a substantial lack of experimental data on the spectroscopic and optical properties of benzene ice and its temperature dependence, especially at Titan-relevant stratospheric conditions. We have therefore analyzed in laboratory the spectral characteristics and evolution of benzene ice's vibrational modes at deposition temperatures ranging from 15 K to 130 K, from the far- to mid-IR spectral region (50 — 8000 cm$^{-1}$). We have determined the amorphous to crystalline phase transition of benzene ice and identified that a complete crystallization is achieved for deposition temperatures between 120 K and 130 K. We have also measured the real and imaginary parts of the ice complex refractive index of benzene ice from 15 K to 130 K. Our experimental results significantly extend the current state of knowledge on the deposition temperature dependence of benzene ice over a broad infrared spectral range, and provide useful new data for the analysis and interpretation of Titan-observed spectra.

---

[1] Corresponding Author Delphine Nna Mvondo: dnnamvo@umbc.edu ; dnnamvondo@gmail.com
[2] University of Maryland Baltimore County (UMBC), Center for Space Sciences and Technology (CSST), 1000 Hilltop Circle, Baltimore, MD 21250, USA
[3] NASA Goddard Space Flight Center, 8800 Greenbelt Road, Greenbelt, MD 20771, USA

## 1. INTRODUCTION

Benzene ($C_6H_6$), the simplest aromatic hydrocarbon, is a molecule that has raised great interests in the astrophysical community for almost four decades. This is mainly because $C_6H_6$ is one of the main precursors of polycyclic aromatic hydrocarbons (PAHs) reported to be present in interstellar dust particles (IDPs) (Leger & Puget 1984; Allamandola et al. 1989; Tielens 2013 and references therein), carbonaceous chondrites (Pering & Ponnamperuma 1971; Hayatsu et al. 1977, 1980; Hahn et al. 1988), and other astrophysical environments, such as carbon-rich, high-temperature environments (circumstellar and carbon-rich protoplanetary nebulae) (Clemett et al. 1994; Buss et al. 1993). Benzene rings easily produces more complex, polycyclic structures by the one-ring build-up mechanism (Simoneit & Fetzer 1996). In space, an analogous process as carbon soot formation occurring on Earth, can be initiated through the completion of that first aromatic ring and may also lead to the synthesis of PAHs (Tielens and Charnley 1997). Mechanisms involving addition of hydrocarbons, such as acetylene onto aromatic rings as well as attachment of other aromatic rings, or hydrocarbon pyrolysis, have been proposed to characterize the growth process of PAHs (Bittner & Howard 1981; Frenklach and Feigelson 1989; Wang & Frenklach 1997; Cherchneff 2011 and references therein). PAHs synthesis from shocked benzene also has been reported (Mimura 1995). PAHs are crucial materials involved in a variety of cosmochemical processes. For example, amino acids could be synthesized by aqueous alteration of precursor PAHs in carbonaceous chondrites (Shock & Schulte 1990). PAHs are also produced in laboratory-simulated planetary atmospheres of Titan and Jupiter (Sagan et al. 1993; Khare et al. 2002, Trainer et al. 2004) and results from these studies indicate that the formation of aromatic rings and polyaromatics may be, among others sources, a possible chemical pathway for the production of the atmospheric solid particles (Lebonnois et al. 2002; Wilson et al. 2003; Trainer et al. 2004). The formation and evolution of benzene in planetary environments or other Solar System objects, thus represents a fundamental primary stage of the PAHs production and other subsequent relevant chemical and prebiotic processes (like soot formation). In this context, several works related to benzene have been devoted to better understand the physico-chemical processes of irradiated $C_6H_6$, in its gaseous and solid phases, and the derived products, by acquiring high-resolution astronomical spectra, carrying out detailed laboratory studies or developing theoretical modelling (Allamandola et al. 1989 and references therein; Callahan et al. 2013; Materese et al. 2015; Mouzay et al. 2021). Laboratory astrophysical investigations have mostly focused on performing vibrational spectroscopy of ion, electron or UV irradiated $C_6H_6$ gas and $C_6H_6$ ice to provide data on the spectral properties of the irradiated $C_6H_6$ materials, compare them with spectra obtained from astronomical observations (e.g. observations of the interstellar medium), or study photo-processed benzene ices to understand the fate of benzene ices in Titan's stratosphere and help understanding the formation of aerosols analogs observed in Saturn moon's stratosphere (Mouzay et al. 2021).

Our experimental study differs from such previous works. The main objective was to carry out a comprehensive experimental study on pristine benzene ice in order to accurately determine its chemical behavior and its infrared (IR) and optical properties at different deposition temperatures. The initial motivation behind our laboratory experiments was to constrain the role of benzene in the formation and chemical composition of Titan's High-Altitude South Polar (HASP) ice cloud. The HASP ice cloud is a massive stratospheric ice cloud system, recently observed in far-IR limb spectra recorded by the Cassini Composite InfraRed Spectrometer (CIRS) in Titan's late southern fall mid stratosphere (~200 km altitude), at high southern polar latitudes (Anderson et al. 2018a).

During the Cassini mission, one of the highly unexpected findings was the detection of benzene gas in Titan's upper atmosphere, as determined by in situ analysis by the Cassini Plasma Spectrometer (CAPS) and the Ion Neutral Mass Spectrometer (INMS) performed during Cassini's first close flybys of Titan (Coates et al. 2007; Waite et al. 2007). During subsequent Titan flybys, $C_6H_6$ gas was also observed by Cassini CIRS in Titan's stratosphere, at high northern and high southern latitudes (Coustenis et al. 2013, 2016, 2018). Atmospheric $C_6H_6$ gas, produced through the photodissociation of gaseous $N_2$—$CH_4$ in Titan's upper atmosphere and successive ion-neutral reactions (Vuitton et al. 2008), enters altitude regions in Titan's stratosphere where it can saturate and condense. Vinatier et al. (2018) reported the contribution of the $\nu4$ vibrational mode of benzene ice with other unidentified molecules during the observation of Titan's south polar stratospheric cloud at mid to upper stratospheric altitudes during southern fall, using CIRS nadir and limb observations in May 2013 and March 2015, respectively. Anderson et al. (2018a) reported that later, in July 2015 during Titan's late southern fall, $C_6H_6$ condenses at altitudes where the HASP cloud vertically resides (Figure 1), deeper in the stratosphere than the $C_6H_6$ ice cloud reported by Vinatier et al. (2018), and the Anderson et al. (2018a) investigators experimentally demonstrated that $C_6H_6$ co-condensed with other organic ice components, likely contributes to the chemical composition of the Titan's HASP ice cloud. The temperatures in Titan's mid to low stratosphere range from ~66 K to 170 K (Schinder et al. 2012, Anderson et al. (2014), Teanby et al. 2019), and experience variations with altitude and season. Given that infrared spectral features depend on the thermal excitation of the crystal's vibrational modes, laboratory infrared spectra of ices measured at various low and high temperatures, including the thermal conditions at different altitudes in Titan's atmosphere, can considerably improve the accuracy of the interpretation of infrared emission spectra obtained from remote sensing observations. Particularly, studying the deposition temperature dependences of benzene ice is necessary to constrain its spectral and optical properties, and ultimately its contribution to Titan's cloudy stratosphere during winter seasons, because the temperature may alter the peak positions, widths, and shapes of the absorption bands of $C_6H_6$ ice.

However, no such thorough study has been performed and reported for benzene ice related to Titan studies. The literature reports few works on the temperature effects on solid benzene (Mair & Hornig 1949; Hollenberg & Dows 1962; Ishii et al. 1996; Mouzay et al. 2021). However, the studies of Mair & Hornig, Hollenberg & Dows and Ishii et al. focused of solid $C_6H_6$ at specific temperatures and did not span a wide range of temperatures that extend from the amorphous to crystalline phases. For example, Mair & Hornig (1949) recorded infrared spectra of $C_6H_6$ ice obtained from the crystallization of the $C_6H_6$ liquid phase at 249 K, 218 K and 103 K from $641 - 5000$ cm$^{-1}$ (15.6 $-$ 2 µm), Hollenberg & Dows (1962) from condensed $C_6H_6$ gas at 85 K and annealed crystalline $C_6H_6$ ice at 155 K from $550 - 1600$ cm$^{-1}$ (18.2 $-$ 6.25 µm), while Ishii et al. (1996) focused on Raman spectroscopy of $C_6H_6$ ice deposited from 17 K to 78 K and the annealed ice from 17 K to 98 K. While the work of Mouzay et al. (2021) reported annealed 16 K-$C_6H_6$ ice from 70 K to 130 K and not $C_6H_6$ ice directly deposited at different temperatures, which is more appropriate to Titan's ice cloud studies, as explained later in Session 2.

In this work, we have conducted a systematic and thorough study of benzene ice with the vapors directly deposited at several temperatures between 15 K and 150 K. Deposition temperatures between 66 K and 130 K are comparable to the condensation temperatures in Titan's mid-to-low stratosphere. We decided to extend the spectral analysis to deposition temperatures lower than 66 K in order to fully understand the behavior of $C_6H_6$ ice with temperature changes, and to precisely identify the transition from its amorphous to crystalline phase in our laboratory sample. Direct vapor deposition studies (as opposed to annealing experiments) are almost nonexistent and are crucial to identify the chemical composition of Titan's CIRS-observed stratospheric ice clouds.

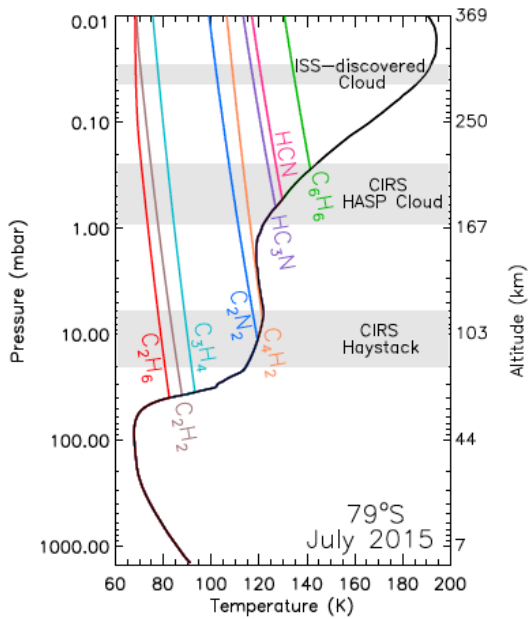

**Figure 1.** Figure 1 from Anderson et al. (2018a). Titan's pressure-temperature-altitude profile (*black curve*) at 79°S during late southern fall (July 2015) and vertical distributions for eight of Titan's stratospheric organic vapors (*color-coded curves*). The intersection between temperature profile and the saturation vapor pressure curves provides the estimated altitude location for the condensation of each individual vapor. The saturation vapor pressure for benzene was taken from Fray & Schmitt (2009).

## 2. EXPERIMENTAL TECHNIQUE

Thin film transmission spectroscopy of amorphous and crystalline phases of pure benzene ice was conducted using the SPECtroscopy of Titan-Related ice AnaLogs (SPECTRAL) ice high-vacuum chamber (Nna-Mvondo et al. 2019; Anderson et al. 2018b), located at NASA Goddard Space Flight Center (GSFC). The technique, used in our laboratory for the IR spectroscopy of ice analogs, consists in depositing the vapor of one or several molecules onto a cold diamond substrate maintained at low temperature (15 K — 200 K) by a close cycle helium cryostat. A thin ice film (≤ 10 μm) forms on the diamond substrate and its spectral properties are monitored in real-time and in situ through a Fourier Transform Infrared (FTIR) spectrometer. The complete description of the SPECTRAL ice chamber, along with the standard experimental methodology for vapor deposition and data analyses utilized in the present work, are detailed in Anderson et al. (2018b), Sections 2, 3 and 4.

$C_6H_6$ ice samples were prepared from pure benzene vapors that were obtained at saturation equilibrium from the purified liquid phase evaporated at room temperature (293 K). At room temperature, benzene has a saturation vapor pressure of 100 mbar. Benzene (Sigma-Aldrich, 99.99% HPLC grade) was purified from air contaminants (predominantly water and carbon dioxide) and other trace organic impurities, by freeze-pump-thaw cycling under vacuum using successive cold baths of ethanol ($C_2H_5OH$) / liquid nitrogen ($LN_2$) (157 K) and acetone ($C_2H_6CO$) / $LN_2$ (179 K). Afterwards, purified $C_6H_6$ was stored for further use in vacuo in an ultra-low temperature freezer at -86°C (187 K).

The thin $C_6H_6$ ice films were obtained by slow deposition of pure benzene gas onto the diamond substrate inside the SPECTRAL chamber, under vacuum ($10^{-8}$ mbar), at a flow rate of 0.66 ml/min, at the desired low temperature, with an ice deposition rate of 0.56 $\mu$m/min. The distance from the injection nozzles to the diamond substrate is ~1.3 cm. Warming and cooling rates of the diamond substrate on which the ice deposits were 4.8 K min$^{-1}$ and 1.3 K min$^{-1}$, respectively. The temperature was controlled with an accuracy of $\pm 0.5$K from 2 K to 305 K by a temperature controller connected to the cryocooler, a 36 Ohm internal resistive Thermofoil heater and two silicon diodes (accuracy $\pm 0.5$ K from 2 K to 305 K), one is located on the holder of the diamond substrate which measure the temperature of the sample, and the other one on the cold finger of the cryocooler (Anderson et al. 2018b). The thicknesses of the ice films at each deposition temperature ($T_d$) were determined by double laser interferometry (detailed in Anderson et al. 2018b, Section 4.3) and using the published $n_0$ value of 1.54 $\pm$ 0.02 at 632.8 nm for benzene ice formed after vapor deposition at 100 K determined by Romanescu et al. (2010). The calculated thickness for $C_6H_6$ ice films  ranged between 3.6 and 4.1 $\mu$m, which corresponds to 26 and 30 $\mu$mol of benzene, respectively.

We decided to obtain the crystalline benzene ice from direct deposition of $C_6H_6$ vapor at low temperature rather than by annealing[1] the sample, a technique commonly used to achieve the crystallization of an ice for most of the planetary ice laboratory studies found in the literature (e.g. Hollenberg & Dows 1962;

---

[1] The annealing process involves depositing the vapor onto the cold substrate in the amorphous phase at very low temperatures (e.g. <100 K), then warming the ice to its appropriate crystalline temperature, and annealing it for several hours at this temperature to allow the molecules to adopt a low-energy configuration. Subsequent cooling/warming of the ice is then achieved to record the infrared absorbance or transmission spectrum at the required temperature.

Khanna et al. 1988; Kim & Kaiser 2009; Zhou et al. 2009; Mouzay et al. 2021). As we demonstrated with $C_2H_5CN$ ice in a recent study by Nna-Mvondo et al. (2019), the annealing experimental method raises important concerns about the quality of the crystalline phase of the annealed ice. Indeed, in this previous work, we revealed spectral differences between the directly-deposited crystalline ice and the annealed ice, indicating that the ice does not reach complete crystallization when annealed, and a portion of the amorphous structure is irreversibly retained in the final crystalline phase. The observed spectral variabilities also affect the optical constants of the ice being studied, i.e. the real ($n$) and imaginary ($k$) parts of the ice complex refractive index (or optical constants), which are needed for radiative transfer fits to observed spectra. As well, the annealing experimental approach, although appropriate to study ices formed in the interstellar medium, is not applicable for simulating ices formed in planetary atmospheres, such as Titan's stratosphere. For example, for Titan, in order to reproduce ice cloud formation in the stratosphere, vapors must be directly deposited at warmer temperatures (e.g., 110 K), since most of Titan's organic vapors condense as the vapors cool while descending throughout Titan's stratosphere (as seen in Figure 1).

The spectra of $C_6H_6$ thin ice films were collected at deposition temperatures ranging from 15 K to 150 K, in the far- to mid-IR spectral region (50 — 8000 $cm^{-1}$; 200 — 1.25 µm). For each resulting transmittance spectrum 256 scans were averaged at a spectral resolution of 4 $cm^{-1}$. The motivation of our work was to use the experimental spectral data that we collected for studying and interpreting the stratospheric HASP ice cloud observed by CIRS in far-IR limb spectra. The spectral dependences of the majority of the CIRS-observed stratospheric ice clouds were determined from analyses of CIRS far IR-targeted low spectral resolution limb scans acquiring many low-resolution spectra, as it was desirable in some circumstances like during observations of strong gas emissions such as methane. Specifically, for the HASP ice cloud observations, the type of far-infrared limb observations was the far-infrared aerosol scan (FIRLMBAER), the second-closest observation to Titan, occurring at (15—25) $10^3$ km (45—75 minutes from closest approach), using a spectral resolution of 15 $cm^{-1}$ (the lowest spectral resolution of CIRS). For this reason, the resolution of 4 $cm^{-1}$ for the collection of our experimental spectra was very convenient and relevant for the objective of our experimental study.

In the mid-IR, the FTIR spectrometer's gain and aperture size were initially optimized to reduce the intensity of the CVD diamond's strong absorption bands, which arise from the sample substrate and IR windows of the

SPECTRAL high-vacuum chamber. This correction is necessary since CVD diamond
strongly absorbs between 2300 $cm^{-1}$ and 2000 $cm^{-1}$. However, between the sub-mm and
near-IR spectral regions, CVD diamond has excellent throughput and using diamond
transmission windows allow for automatic transitioning across the far- and mid-
IR spectral regions while keeping the purge and experimental conditions intact.
Thus, for each temperature, far and mid-IR spectra can be recorded consecutively
in the exact same experimental conditions. Prior to vapor deposition, background
spectra were collected for each selected temperature. Ice absorption spectra
were then recorded and automatically background corrected, i.e. the single-beam
spectrum I of the ice sample deposited on the window was ratioed against the
single-beam spectrum Io obtained with the chamber IR windows but without the
ice sample. This background correction also allowed to remove spectral features
introduced by the spectrometer and the remaining traces of ambient atmospheric
gases (while continuously purging the IR spectrometer compartments with dried
compressed air). The absorption spectrum was then converted to absorbance.

Each laboratory-acquired IR absorbance spectrum shown in this paper was
corrected to remove the channel fringes that are produced by the reflections at
the vacuum-ice and ice-substrate sample surface (Hirschfeld and A. W. Mantz
1976). When removing the channel fringes from the spectrum, the absorbance value
was purposely "zeroed in" in the far-IR region for frequencies below 100 $cm^{-1}$
when the signal-to-noise (S/N) becomes low (S/N < 3) and the peaks cannot be
accounted as real absorption bands. Due to the high noise level at the low
wavenumber end between 50 $cm^{-1}$ and 100 $cm^{-1}$, all wavenumbers below 100 $cm^{-1}$ were
then set at the 100 $cm^{-1}$ value for generating the corrected absorbance spectra
as well as for computing the optical constants $n$ and $k$.

After the channel fringes were removed, the corresponding $n$ and $k$ values
were computed for each corrected absorbance spectrum. The detailed procedure
for computing the optical constants was described previously in Anderson et al.
(2018b) in Section 4.3. In summary, we monitored the $C_6H_6$ thin ice films as they
grew with deposition time via double laser interferometry and then we used
Braggs Equation to compute the thickness $d$ (see e.g., Tempelmeyer and Mills,
Jr., 1968; Domingo et al., 2007); an $n_0$ of 1.54 at 632.8 nm was adopted from
Romanescu et al., (2010). We then estimated the imaginary part of the refractive
index ($k$) from the relation $k(\nu) = \alpha(\nu)/4\pi\nu$, where the Lambert absorption
coefficient $\alpha(\nu)$ was initially estimated from the relation $-(ln\ T)/d$; $T$ is the
ice sample fringe-corrected transmittance spectrum, $d$ is the total ice
thickness, and $\nu$ is wavenumber in $cm^{-1}$. Once the estimated $k$ values were computed,
we then employed the Kramers-Kronig relationship to calculate the initial
estimate of the real part of the refractive index ($n$) as a function of wavenumber. The initial estimates of both $n(\nu)$ and $k(\nu)$ were then input into the fully expanded Lambert absorption coefficient equation to compute $\alpha(\nu)$ (see also Rocha & Pilling, 2014). $\alpha(\nu)$ was then input back into $k(\nu) = \alpha(\nu)/4\pi\nu$ to compute the new $k(\nu)$ values, and we continued to iteratively adjust the values of $k(\nu)$ to compute the new $n(\nu)$ values until the $n(\nu)$ and $k(\nu)$ values converged to their final states.

All the original data files of the optical constants as well as the absorbance spectra obtained in this study for $C_6H_6$ ice at the different deposition temperatures will be made available on the NASA website https://science.gsfc.nasa.gov/691/spicelab.

## 3. EXPERIMENTAL $C_6H_6$ ICE SPECTRA

Benzene ice was studied from the amorphous to the crystalline phase and we examined the variation of the $C_6H_6$ ice absorption bands, from the far- to mid-IR spectral region, at the deposition temperatures ranging from 15 K to 150 K. This study was performed in an effort to better understand the $C_6H_6$ ice phase transitions with the corresponding observed spectral changes, and to identify when the complete crystallization of benzene ice is achieved. Specifically, we have recorded absorbance spectra at the deposition temperatures of 15 K, 30 K, 32 K, 35 K, 45 K, 60 K, 90 K, 100 K, 110 K, 120 K, 125 K, 130 K, 135 K, 140 K, and 150 K (Figures 2 and 3). We have also studied the temporal variations of the $C_6H_6$ ice absorption bands, i.e. the time evolution of the $C_6H_6$ ice spectra after vapor deposition for each deposition temperature cited earlier. Contrary to our previous results on propionitrile ice (Nna-Mvondo et al. 2019), we did not observe any temporal variation of the $C_6H_6$ ice band positions, nor any variation in the band intensities and shapes for any of the $C_6H_6$ absorption bands at any given temperature, even 24 hours after deposition. Therefore, in this paper we only present the results for $C_6H_6$ ice spectra obtained at different deposition temperatures.

The absorption bands of $C_6H_6$ ice were assigned in the full far- to mid-IR spectral range studied in this work ($50 - 8000$ cm$^{-1}$; $200 - 1.25$ μm), based on previous spectral studies of benzene in its liquid, gas and solid phase (Mair & Hornig 1949; Bertie & Keefe 2004; Miani et al. 2000). Over the last few decades, experimental and theoretical investigations of the vibration modes of benzene have become increasingly complete due to refined laboratory techniques and increased computational power. The vibrational modes of benzene have been extensively studied so that we were able to assign most of them in our experimental $C_6H_6$ ice IR spectra. All the vibrational assignments observed in the $C_6H_6$ ice IR spectra that we recorded in our laboratory are listed in Table 1, for the amorphous and crystalline phases of benzene ice, deposited at 15 K and 130 K, respectively. The vibrational modes assignments are numbered according to Herzberg's notation (Herzberg 1945). We have considered with confidence spectral peaks in the infrared spectra with signals rising above the $3\sigma$ noise level as $C_6H_6$ ice absorption bands. Spectral features with a low signal-to-noise ratio (S/N < 3) were disregarded. Apart from the assignment of the $2\nu_{CH}$ first overtone mode at 5980 cm$^{-1}$ (at 15 K), it was not possible to assign all other spectral features above 4600 cm$^{-1}$ because at these higher wavenumbers anharmonic effects are so large that the local-mode (LM) analysis is required (Bertie & Keefe 2004; Wyatt 1998). The LM analysis technique is usually developed to describe multiple-quantum (overtone) transitions where vibrational energy tends to become localized on a single bond. It differs from the normal-mode (NM) analysis of vibrations in polyatomic molecules that describes single-quantum transitions.

Benzene molecule belongs to the point group $D_{6h}$. It has a six-fold axis of symmetry ($C_6$) along a line through the center of the molecule and perpendicular to the plane of the paper, six vertical planes of symmetry ($\sigma_v$) through $C_6$ at angles of 30° to one another, and one horizontal plane of symmetry ($\sigma_h$) perpendicular to $C_6$. $C_6H_6$ has twelve different symmetry types (species) ($A_{1g}$, $A_{1u}$, $A_{2g}$, $A_{2u}$, $B_{1g}$, $B_{1u}$, $B_{2g}$, $B_{2u}$, $E_{1g}$, $E_{1u}$, $E_{2g}$ and $E_{2u}$) and twenty fundamental vibrations (see Table 1), which were all detected in the $C_6H_6$ ice spectra that we recorded from 15 K to 130 K: $\nu_1(A_{1g})$, $\nu_2(A_{1g})$, $\nu_3(A_{2g})$, $\nu_4(A_{2u})$, $\nu_5(B_{1u})$, $\nu_6(B_{1u})$, $\nu_7(B_{2g})$, $\nu_8(B_{2g})$, $\nu_9(B_{2u})$, $\nu_{10}(B_{2u})$, $\nu_{11}(E_{1g})$, $\nu_{12}(E_{1u})$, $\nu_{13}(E_{1u})$, $\nu_{14}(E_{1u})$, $\nu_{15}(E_{2g})$, $\nu_{16}(E_{2g})$, $\nu_{17}(E_{2g})$, $\nu_{18}(E_{2g})$, $\nu_{19}(E_{2u})$, and $\nu_{20}(E_{2u})$.

**Table 1**

Infrared vibration modes and frequencies of amorphous and crystalline $C_6H_6$
ices from the laboratory spectra obtained in this study

| Amorphous (15 K) IR Frequencies ($cm^{-1}$)* | Crystalline (130 K) IR Frequencies ($cm^{-1}$)* | Band Assignment[#] | Ref.[§] |
|---|---|---|---|
| 405 *vw* | 404 *vw* | $\nu_{20}$ ring deform. | *b, c* |
|  | 418 *vw* |  |  |
| 608 *vw*[‡] | 611 *vw*[‡] | $\nu_{18}$ ring deform. | *c* |
| 619 *vw*[‡] | 621 *vw*[‡] |  |  |
| 679 *vs* | 681 *vs* | $\nu_4$ C–H asym. bending | *a, b, c* |
| 703 *vw* | 706 *m* | $\nu_8$ ring deform. | *a, b, c* |
| S/N < 3 | 785 *vvw* | $\nu_{17} - \nu_{20}$ | *b* |
| 854 *w* | — | $\nu_{11}$ C–H asym. bending | *a, b, c* |
| 973 *w* | 975 *w* | $\nu_{19}$ C–H asym. bending | *a, b, c* |
|  | 986 *w* |  |  |
| 991 *vw* | — | $\nu_7$ C–H asym. bending | *b, c* |
| 999 *vw* | — | $\nu_2$ C–C stretching | *a, c* |
| 1011 *vw* | 1009 *vw* | $\nu_6$ C–C sym. bending | *a, b, c* |
| 1036 *s* | 1033 *s*[‡] | $\nu_{14}$ C–H sym. bending | *a, b, c* |
|  | 1038 *s*[‡] |  |  |
| S/N < 3 | 1109 *vw*[‡] | $\nu_8 + \nu_{20}$ | *a, b* |
|  | 1119 *vw*[‡] |  |  |
| 1146 *w* | 1143 *w*[‡] | $\nu_{10}$ C–H sym. bending | *a, b, c* |
|  | 1148 *w*[‡] |  |  |
| 1176 *w* | — | $\nu_{17}$ C–H sym. bending | *a, b, c* |
| 1243 *vw* | 1250 *vw* | $\nu_{10} + \nu_{16}$ | *a, b* |
| 1256 *vw* | 1259 *vw* |  |  |
|  | 1268 *vw* |  |  |
|  | 1277 *vw* |  |  |
| 1311 *vw* | 1312 *vw* | $\nu_9$ C–C stretching | *a, b, c* |
| 1348 *vw* | — | $\nu_3$ C–H sym. bending | *a, b, c* |
| 1399 *vw* | 1402 *vw*[‡] <br> 1413 *vw*[‡] | $\nu_7 + \nu_{20}$ | *a, b* |
| 1478 *vs* | 1478 *vs* | $\nu_{13}$ C–C stretching | *a, b, c* |
| 1539 *w* | 1549 *w*[‡] <br> 1561 *w*[‡] | $\nu_4 + \nu_{11}$ | *a, b* |
| 1586 *vw* | — | $\nu_{16}$ C–C stretching | *a, b, c* |
| 1604 *vw* | — | $\nu_2 + \nu_{18}$ | *a, b* |
| 1617 *vw* | 1617 *vw* | $\nu_6 + \nu_{18}$ | *a, b* |
| S/N < 3 | 1644 *vw* | $\nu_{14} + \nu_{18}$ | *a, b* |
| 1672 *vw* | 1679 *vw*[‡] <br> 1684 *vw*[‡] | $\nu_8 + \nu_{19}$ | *a, b* |
| 1713 *vw* | 1714 *vw* | $\nu_6 + \nu_8$ | *a, b* |
| 1754 *vw* | 1754 *vw* | $\nu_{10} + \nu_{18}$ | *a, b* |
| 1823 *m* | 1829 *m*[‡] <br> 1838 m[‡] | $\nu_{11} + \nu_{19}$ | *a, b* |
| 1966 *w* | 1976 *w*[‡] <br> 1981 *w*[‡] | $\nu_7 + \nu_{19}$ | *a, b* |
| 2326 *vw* | 2325 *vw* | $\nu_{10} + \nu_{17}$ | *a, b* |
| 2383 *vw* | 2381 *vw* | $\nu_3 + \nu_{14}$ | *a, b* |
| S/N < 3 | 2466 *vw* <br> 2482 *vw* <br> 2490 *vw* | $\nu_9 + \nu_{17}$ | *b* |
| 2594 *vw* | 2592 *vw* | $\nu_6 + \nu_{16}$ | *a, b* |
| 2613 *vw* | 2612 *vw* | $\nu_6 + (\nu_{16} + \nu_{18})$ | *b* |
| 2651 *vw* | 2648 *vw* | $\nu_{13} + \nu_{17}$ | *a, b* |
| — | 2728 *vw* | $\nu_{10} + \nu_{16}$ | *b* |
| 2783 *vw* | — | $\nu_{10} + (\nu_2 + \nu_{18})$ | *b* |
| 2819 *vw* | 2818 *vw* | $\nu_3 + \nu_{13}$ | *a, b* |
| | 2831 *vw* | | |
| — | 2849 *vw* | Not assigned | |
| 2887 *vw* | 2889 *vw* | Not assigned | |
| 2907 *vw* | 2909 *vw* | $\nu_9 + \nu_{16}$ | *a, b* |
| 2926 *vw* | 2933 *vw* | | |
| 2947 *vw* | 2944 *vw* | | |
| | 2953 *vw* | | |
| 3004 *vw* | 3005 *vw* | Not assigned | |
| 3033 *s* | 3030 *s*‡ | $\nu_{12}$ C-H stretching | *a, b, c* |
| | 3037 *s*‡ | | |
| 3070 *m* | 3068 *m* | $\nu_5, \nu_{15}$ C-H stretching | *a, c* |
| 3089 *s* | 3086 *s* | $\nu_1$ C-H stretch., $\nu_{13} + \nu_{16}$ | *a, b* |
| 3447 *vw* | 3457 *vw* | $\nu_{15} + \nu_{20}$ | *a, b* |
| 3610 *vw* | 3610 *vw* | $\nu_5 + \nu_{18}$ | *a, b* |
| 3639 *vw* | 3635 *vw*‡ | $\nu_{12} + \nu_{18}$ | *a, b* |
| | 3641 *vw*‡ | | |
| S/N < 3 | 3659 *vw* | Not assigned | |
| 3684 *vw* | 3682 *vw*‡ | $(\nu_2 + \nu_{13} + \nu_{18}) + \nu_{18},$ | *b* |
| 3694 shoulder | 3691 *vw*‡ | $(\nu_{13} + \nu_{16}) + \nu_{18}$ | *b* |
| 3934 *vw* | 3931 *vw* | $\nu_{11} + \nu_{12}$ | *a* |
| 3955 *vw* | 3953 *vw* | $\nu_3 + \nu_6 + (\nu_2 + \nu_{18})$ | *b* |
| 4021 *vw* | 4017 *vw*‡ | $\nu_2 + \nu_{12}$ | *b* |
| | 4023 *vw*‡ | | |
| — | 4032 | Not assigned | |
| 4056 *m* | 4052 *m* | $\nu_6 + \nu_{15}$ | *b* |
| 4079 *vw* | 4073 *vw*‡ | $\nu_{14} + \nu_{15}$ | *b* |
| | 4080 *vw*‡ | | |
| S/N < 3 | 4099 *vw* | $\nu_1 + \nu_{14}$ | *b* |
| S/N < 3 | 4156 *vw* | $\nu_{10} + \nu_{15}$ | *b* |
| 4174 *vw* | 4168 *vw*$^{\sharp}$<br>4174 *vw*$^{\sharp}$ | $\nu_{10} + \nu_{15}$ | *a, b* |
| 4193 *vw* | 4189 *vw* | $\nu_{12} + \nu_{17}$ | *b* |
| S/N < 3 | 4203 *vw* | Not assigned | |
| S/N < 3 | 4237 *vw* | $\nu_5 + \nu_{17}$ | *a, b* |
| 4258 *vw* | 4256 *vw* | $\nu_{17} + (\nu_{13} + \nu_{16})$ | *b* |
| 4348 *vw* | 4347 *vw* | Not assigned | |
| 4375 *vw* | 4361 *vw*<br>4372 *vw*$^{\sharp}$<br>4377 *vw*$^{\sharp}$ | $\nu_9 + \nu_{15}, \ \nu_3 + \nu_{12}$ | *a, b* |
| 4411 *vw* | 4409 *vw* | $\nu_3 + (\nu_2 + \nu_{13} + \nu_{18})$ | *b* |
| 4429 *vw* | 4423 *vw*$^{\sharp}$<br>4428 *vw*$^{\sharp}$ | $\nu_3 + (\nu_{13} + \nu_{16})$ | *b* |
| S/N < 3 | 4481 *vvw* | Not assigned | |
| 4541 *vw* | 4520 *vvw*$^{\sharp}$<br>4529 *vw*$^{\sharp}$<br>4541 *vw*$^{\sharp}$ | $\nu_{13} + \nu_{15}$ | *b* |
| 4581 *vw* | 4580 *vw* | $\nu_1 + \nu_{13}$ | *b* |
| 4602 *vw* | 4602 *vw*$^{\sharp}$<br>4609 *vw*$^{\sharp}$ | Not assigned | |
| 4619 *vw* | 4618 *vw* | Not assigned | |
| 4641 *vw* | 4640 *vw* | Not assigned | |
| 4667 *vw* | 4663 *vw*$^{\sharp}$<br>4673 *vw*$^{\sharp}$ | Not assigned | |
| 4686 *vw* | 4684 *vw* | Not assigned | |
| 5326 *vvw* | 5345 *vvw* | | |
| S/N < 3 | 5513 *vvw* | Not assigned | |
| S/N < 3 | 5530 *vvw* | Not assigned | |
| S/N < 3 | 5542 *vvw* | Not assigned | |
| S/N < 3 | 5605 *vw* | Not assigned | |
| 5637 *vvw* | 5636 *vw* | Not assigned | |
| 5689 *vw* | 5680 *vvw*[‡] | Not assigned | |
| | 5687 *vw*[‡] | | |
| | 5692 *vw*[‡] | | |
| | 5707 *vw*[‡] | | |
| 5773 *vw* | 5768 *vw*[‡] | Not assigned | |
| | 5777 *vw*[‡] | | |
| S/N < 3 | 5858 *vw*[‡] | Not assigned | |
| | 5864 *vw*[‡] | | |
| S/N < 3 | 5877 *vw* | Not assigned | |
| 5910 *vw* | 5909 *vw* | Not assigned | |
| S/N < 3 | 5926 *vw* | Not assigned | |
| S/N < 3 | 5947 *vw* | Not assigned | |
| S/N < 3 | 5961 *vw* | | |
| 5980 *vw* | 5976 *vw*[‡] | $2\nu_{CH}$ 1[st] overtone | *b* |
| | 5989 *vw*[‡] | (C–H stretch) | |
| | 5997 *vw*[‡] | | |
| | 6019 *vw*[‡] | | |
| 6066 *vw* | 6047 *vw*[‡] | Not assigned | |
| | 6061 *vw*[‡] | Not assigned | |
| | 6074 *vw*[‡] | Not assigned | |
| 6140 *vw* | 6135 *vw* | Not assigned | |
| 6178 *vw* | 6171 *vw*[‡] | Not assigned | |
| | 6184 *vw*[‡] | Not assigned | |

**Note.** [a] Intensities of band: *vs* very strong, *s* strong, *m* medium, *w* weak, *vw* very weak, *vvw* very very weak. [‡] Multiple frequencies appear for the same vibrational assignment designate a band that is split into two, three or four. Features for which frequencies are not indicated, are below three times the noise level (S/N < 3) and therefore are not considered with sufficient confidence as absorption bands of $C_6H_6$ ice. — Non-observed frequencies. [#] The fundamental vibrations are numbered following Herzberg (1945). [§] The band assignments are based on

crystalline ice spectral data published in (*a*) Mair & Hornig (1949) and on liquid and gas phase data reported in (*b*) Bertie & Keefe (2004) and (*c*) Miani et al (2000), respectively. Bertie & Keefe (2004) uses Herzberg notation, while Mair & Hornig (1949) and Miani et al (2000) follow the notation of Wilson (1934 and 1955).

In the next sections, we describe and analyze in detail the IR absorbance spectra of benzene ice obtained in this study[2] and their evolution with temperature, in the far-IR region from 50 $cm^{-1}$ to 640 $cm^{-1}$ (100 — 15.6 μm), Section 3.1, and in the Mid-IR region from 640 $cm^{-1}$ to 8000 $cm^{-1}$ (15.6 — 1.25 μm), Section 3.2.

### 3.1. *Far-IR $C_6H_6$ Ice Spectra (50 $cm^{-1}$ to 640 $cm^{-1}$)*

Figure 2 shows $C_6H_6$ ice far-IR spectra obtained in our laboratory for benzene vapors deposited at temperatures between 15 K and 130 K. Even though we carried out experiments at higher temperature (135 K, 140 K, and 150 K), we observed that at temperatures higher than 130 K, $C_6H_6$ ice sublimes under the experimental conditions of the SPECTRAL chamber, in which a vacuum pressure is constantly maintained at $10^{-8}$ mbar during the vapor deposition, and the recording of the IR spectra. Thus, in this article we present the experimental results for $T_d \leq 130$ K.

The far-IR spectral region between 50 - 400 $cm^{-1}$ (100 - 25 μm), did not display any absorption bands of benzene ice defined as detectable (spectral features with a signal-to-noise ratio S/N > 3), for any of the deposition temperatures at which the spectra were collected. In this low-frequency range, several absorption bands between 55 $cm^{-1}$ and 130 $cm^{-1}$ have been previously assigned to vibrations characteristic of the crystal lattice. These torsional lattice vibrations, which correspond to out-of-plane and in-phase librations of all benzene molecules about the three crystal axes, are infrared- and Raman-active. They have been previously observed in Raman spectra of solid benzene (Epstein & Steiner 1934; Bonadeo et al. 1972; Ishii et al. 1996), as well in the far-IR spectra of crystalline $C_6H_6$ at 173 K (Chantry et al. 1967), 138 K (Harada & Shimanouchi 1967), 100 K Harada & Shimanouchi 1971), and from 15 K to 150 K (Sataty et al. 1973, Sataty & Ron 1976). We did not identify any of these absorption bands in our experimental far-IR spectra, even for thicker ice films (10 μm), since the signal in this low-energy far-IR spectral region has a low spectral resolution (due the intense noise produced by our FTIR spectrometer below 100 $cm^{-1}$). Therefore our experimental system and procedure did not allow to distinguish them. Bertie & Keefe (2004) reported very weak bands at 264 $cm^{-1}$ and 301.6 $cm^{-1}$ for liquid benzene at 25 ºC. With our instrument and experimental protocol, we are able to detect absorption bands at this frequency range (with absorbance height as low as 0.002 absorbance unit for S/N>3); for example we detected the weak far-IR absorption bands of propionitrile between 100 $cm^{-1}$ and 390 $cm^{-1}$ (Anderson et al., 2018; Nna-Mvondo et al. 2019). This was not the case for C6H6 ice. Very possibly, the bands of liquid $C_6H_6$ measured between 100 and 300 $cm^{-1}$ are very weak bands with intensity below the noise level of our spectra.

---

[2] The raw spectra were recorded with the software of the FTIR spectrometer sets for collecting data at 4 $cm^{-1}$ resolution, however the default/standard configuration for normal work for a resolution of 4 $cm^{-1}$, with a zero-filling factor of 2, sets the data spacing automatically at 2 $cm^{-1}$, i.e. the spectra are automatically oversampled by a factor of 2 in the standard configuration. The IR spectra displayed in figures, are all shown in the default configuration as collected initially, i.e. with wavenumber spaced by 2 $cm^{-1}$ (oversampled by 2).

At frequencies higher than 400 $cm^{-1}$, two vibration modes of $C_6H_6$ ice, the $v_{20}$ and $v_{18}$ ring deformation bands, were detected as shown in Figure 2 and listed in Table 1. For both vibrational modes, we observed spectral variations with the deposition temperature involving significant changes in absorption band position, intensity, and shape (Figure 2).

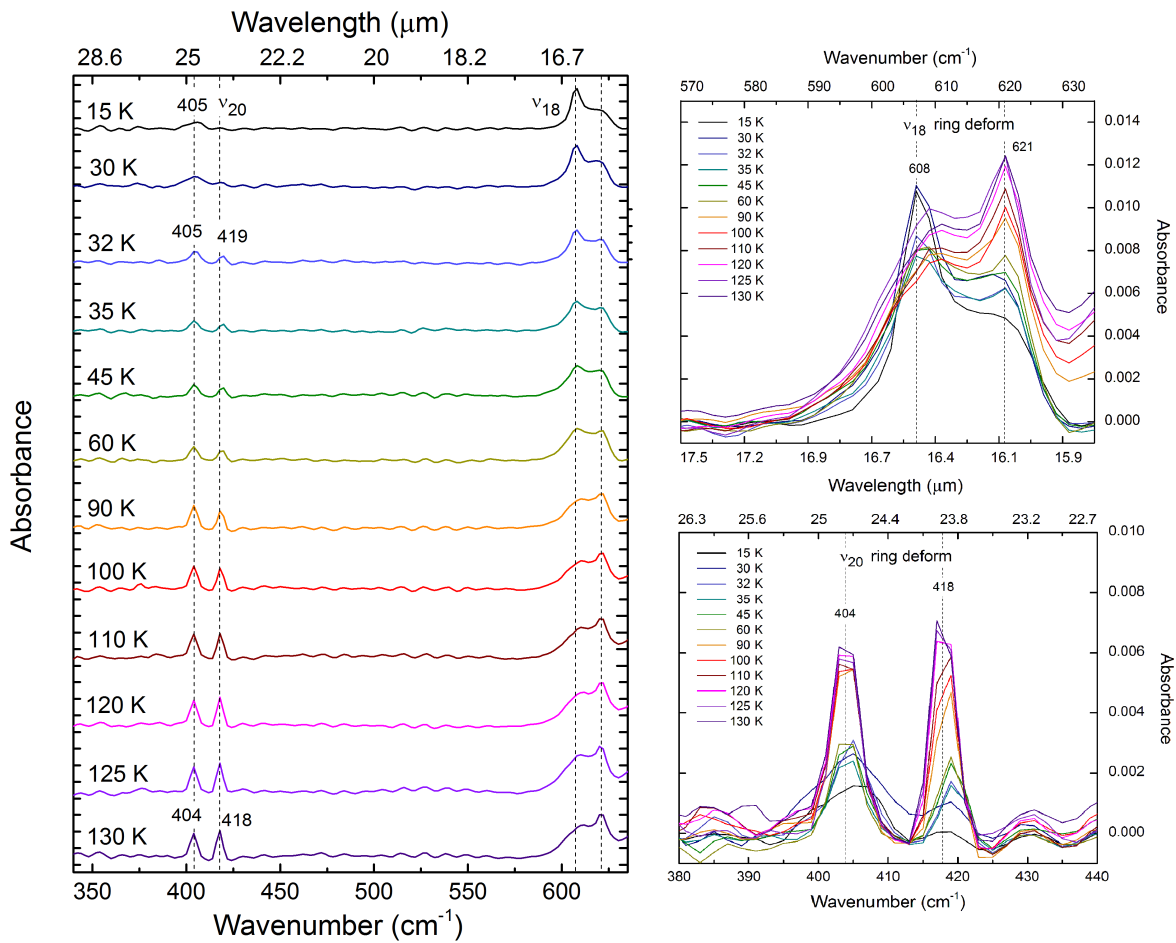

**Figure 2.** Far-IR absorbance spectra of $C_6H_6$ ice (4 $\mu m$ thick film) obtained after deposition of the benzene vapors at temperatures from 15 K to 130 K. Thicknesses of ice films at 15 K, 30 K, 32 K, 35 K, 45 K, 60K, 90 K, 100 K, 110 K, 120 K, 125 K and 130 K were 3.87 $\mu m$, 3.82 $\mu m$, 3.92 $\mu m$, 3.70 $\mu m$, 4.08 $\mu m$, 3.70 $\mu m$, 3.71 $\mu m$, 3.59 $\mu m$, 3.63 $\mu m$, 3.42 $\mu m$, 3.69 $\mu m$, 3.63 $\mu m$, respectively. The left panel shows the entire absorbance spectra from 340 $cm^{-1}$ to 635 $cm^{-1}$ (29.4 − 15.7 $\mu m$). The panels on the right depict specific spectral regions of $C_6H_6$ ring deformation modes, the $\nu_{18}$ band (in the upper panel) and the $\nu_{20}$ band (in the lower panel). Superimposed in the figures are the vibrational transitions and the numbers are peak frequencies in $cm^{-1}$ for the observed ice absorption bands.

At 15 K and 30 K, the $\nu_{20}$ ring deformation mode peaks at 405 $cm^{-1}$ and 404 $cm^{-1}$, respectively, and appears as a broad band. From 32 K and higher, this band is split into an asymmetrical doublet. At 32 K, the $\nu_{20}$ doublet-band branches at 405 $cm^{-1}$ and 419 $cm^{-1}$, with the first maximum at 405 $cm^{-1}$ more intense than the second maximum at 419 cm$^{-1}$. With increasing deposition temperature, the $v_{20}$ doublet increases in intensity and in the splitting, and shifts slightly to lower energies (Figure 2). At 110 K, the doublet is symmetrical, with an equalization of the intensities of the doublet, but from 120 K and higher, the two maxima reverse in intensity with the second maximum more intense than the first one. At 130 K, the doublet is shifted by 1 cm$^{-1}$ and peaks at 404 cm$^{-1}$ and 418 cm$^{-1}$.

The $v_{18}$ ring deformation mode of $C_6H_6$ ice also undergoes significant spectral changes with the deposition temperature (Figure 2). The band appears as a doublet with a main maximum and an inflection at all deposition temperatures, but poorly resolved compared to the $v_{20}$ absorption band that appears with two distinct maxima. At 15 K, the main maximum peaks at about 608 cm$^{-1}$ and the inflection at about 615 cm$^{-1}$. With increasing deposition temperature, the maximum of the main band weakens and displaces towards higher frequencies. At 90 K, the maximum is relocated at 621 cm$^{-1}$, thus undergoes a shift of 13 cm$^{-1}$ to higher energies. No further shift of the maximum is observed at higher deposition temperatures, but its intensity at 621 cm$^{-1}$ increases with the temperature.

### 3.2. *Mid-IR $C_6H_6$ Ice Spectra (640 cm$^{-1}$ to 8000 cm$^{-1}$)*

Figure 3 displays the mid-IR spectra of benzene ice that we have obtained for deposition temperatures ranging from 15 K to 130 K.

Just like the liquid and gas phases, the mid-IR spectra of $C_6H_6$ ice (Figure 3) depicts four C–H stretching modes ($v_1$, $v_5$, $v_{12}$ and $v_{15}$), four C–C stretching modes ($v_2$, $v_9$, $v_{13}$, $v_{16}$), four C–H symmetric (in-plane) bending modes ($v_3$, $v_{10}$, $v_{14}$, $v_{17}$), four C–H asymmetric (out-of-plane) bending modes ($v_4$, $v_7$, $v_{11}$, $v_{19}$) and one C–C symmetric bending mode ($v_6$), among the 30 molecular normal modes of vibration of benzene that produces twenty different frequencies (see Table 1). Most of the fundamental vibrations of benzene ice are observed between 650 cm$^{-1}$ and 1600 cm$^{-1}$ (Figures 3–5). From 1600 cm$^{-1}$ to 6200 cm$^{-1}$, the $C_6H_6$ ice spectrum is dominated by weak combination bands and only the four fundamental C–H stretching vibration modes $v_1$, $v_5$, $v_{12}$ and $v_{15}$ are detected between 3000 cm$^{-1}$ and 3090 cm$^{-1}$ (Figures 3 and 4). In the spectral region 6200 − 8000 cm$^{-1}$ (1.61 − 1.25 μm), for ice film thicknesses ranging from 3.6 to 4.1 μm, benzene ice does not show any absorption bands (Figure 3).

We mention that there were no $C_6H_6$ ice absorption bands detected between 1940 cm$^{-1}$ and 2300 cm$^{-1}$ due to the intense absorption of the diamond, which is the optical material of the sample substrate and IR windows. Even when the parameters of the FTIR spectrometer were optimized to reduce the intensity of the diamond bands, the mid-IR region between 1940 cm$^{-1}$ and 2300 cm$^{-1}$ still showed intense absorption bands of the diamond. As a result, in the fringe-corrected mid-IR spectra shown in Figure 3, the diamond absorption bands were removed by averaging the absorbance values around 1860 cm$^{-1}$ and 1950 cm$^{-1}$ and maintaining this averaged value constant at all wavenumbers in between. Similarly, this approach was applied to the wavenumbers between 2000 cm$^{-1}$ and 2300 cm$^{-1}$.

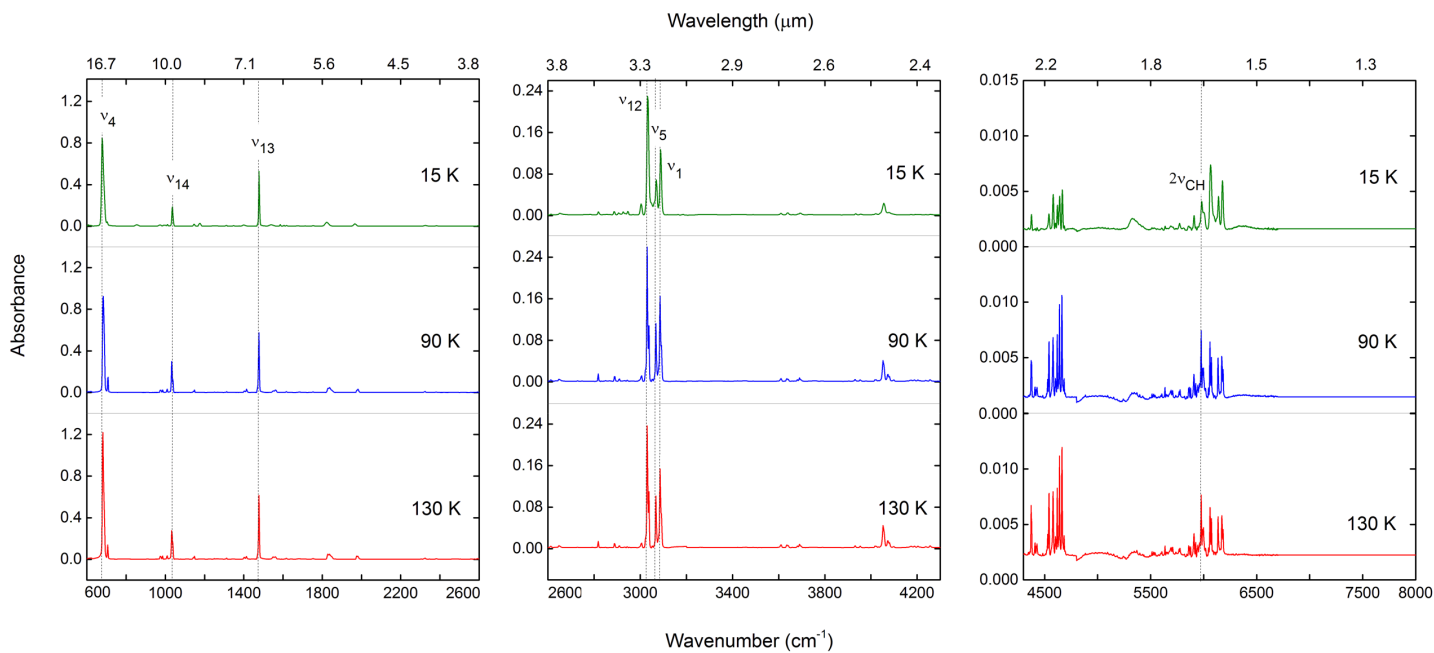

**Figure 3.** Full-range mid-IR absorbance spectra of $C_6H_6$ ice (4 $\mu$m thick film) obtained after deposition of the benzene vapors at temperatures of 15 K, 90 K and 130 K (thin ice films of 3.87 $\mu$m, 3.71 $\mu$m, and 3.63 $\mu$m, respectively).

As with the far-IR region, the mid-IR spectra of $C_6H_6$ ice revealed spectral variations with deposition temperature and significant changes in the position, intensity, and shape of the absorption bands, as displayed in Figures 4—6.

Figure 4 represents the spectral regions in the mid-IR of the most intense fundamental vibration modes of benzene ice and their evolution with the deposition temperature. Every fundamental mode revealed a different spectral behavior with temperature. The $\nu_4$ C–H out-of-plane bending mode (wag vibration), which is the most intense vibrational band of $C_6H_6$ ice, peaks at 679 cm$^{-1}$ at 15

K. As the temperature increases, it becomes sharper (the FWHM[3] is reduced by 4 $cm^{-1}$ from 15 K to 130 K) with an increase in band intensity and shifts to higher energies. At 125 K, the position and intensity of the band stabilizes and is shifted by 2 $cm^{-1}$ peaking at 681 $cm^{-1}$ at 125 K and 130 K. Surprisingly, the largest shift toward highest frequencies is not observed at the highest temperature but at 60 K and 90 K, with a displacement of 5 $cm^{-1}$ from the band position at 15 K. The weak $\nu_8$ ring deformation mode at 703 $cm^{-1}$ at 15 K undergoes a similar spectral evolution with temperature as with the $\nu_4$ vibrational band, although the largest spectral shift occurs at lower temperatures between 35 K and 60 K, and the band reaches a maximum intensity at 90 K and 100 K. The intense $\nu_{13}$ C–C stretching mode of $C_6H_6$ ice at 1478 $cm^{-1}$ does not undergo significant spectral changes with temperature, but only a slight increase in band intensity and, at 90 K and 100 K, a small shift of about 0.5 $cm^{-1}$ towards lower energies are observed. As well, a small shoulder around 1470 $cm^{-1}$ is growing with increasing temperature. One possible explanation of this shoulder is that at 4 $cm^{-1}$ spectral resolution some of the vibrational bands are not fully resolved and some side peaks may appear as shoulder if positioned at less than the spectral resolution from the main band. The $\nu_{14}$ C–H in-plane bending mode is another intense absorption band of $C_6H_6$ ice. It peaks as a singlet band at 1036 $cm^{-1}$ at 15 K and starts to split in two at 32 K to become an asymmetric doublet at 60 K. From 125 K, the doublet is stable and has an intense maximum at 1033 $cm^{-1}$ and a second weaker maximum at 1038 $cm^{-1}$. The four C–H stretching modes $\nu_1$, $\nu_5$, $\nu_{12}$ and $\nu_{15}$ of $C_6H_6$ ice observed between 3000 − 3090 $cm^{-1}$ evolve distinctly with deposition temperature. The $\nu_{12}$ mode at 3033 $cm^{-1}$ at 15 K has a similar behavior as the $\nu_{14}$ band. It splits into two maxima as the temperature increases, which stabilize from 125 K and form an asymmetric doublet peaking at an intense maximum at 3031 $cm^{-1}$, and another one at 3037 $cm^{-1}$. Like band, a growing shoulder with increasing temperatures is observed at higher energy around 3023 $cm^{-1}$. The main spectral changes for the $\nu_1$, $\nu_5$ and $\nu_{15}$ absorption bands are a shift to lower energies of 3 $cm^{-1}$ and 2 $cm^{-1}$, respectively, from 15 K to 130 K and an increase in the band intensity. A small inflection starts to appear in the $\nu_1$ stretch band at 45 K but never achieves a resolved doublet with maxima. As for the $\nu_{12}$ vibration mode, from 125 K, no more spectral changes are observed for the $\nu_1$, $\nu_5$ and $\nu_{15}$ stretch bands.

---

[3] Full Width at Half Maximum

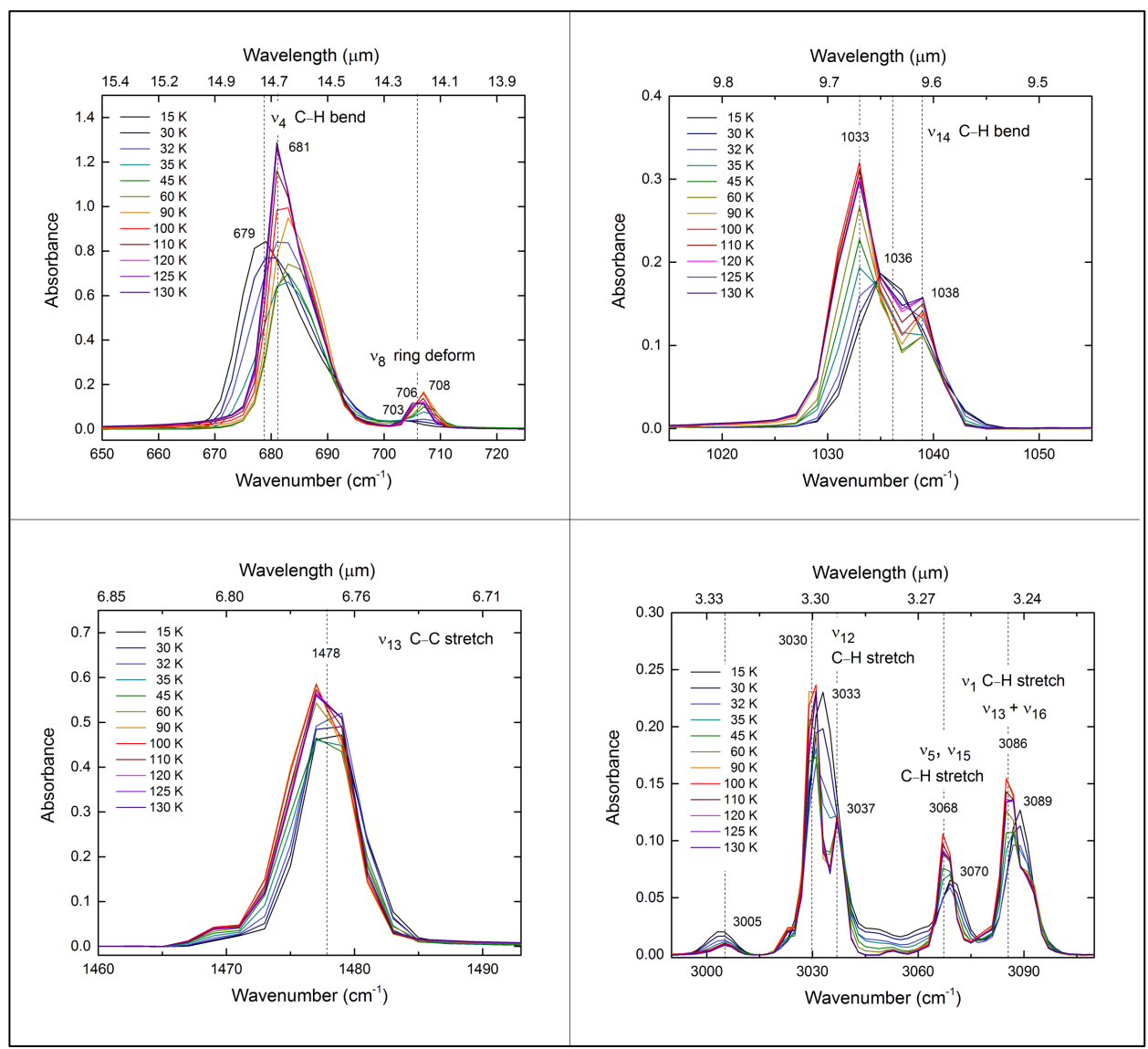

**Figure 4.** Most intense fundamental modes of $C_6H_6$ ice in the mid-IR region from absorbance spectra obtained after deposition of the benzene vapors at temperatures from 15 K to 130 K. Ice film thickness ranges between 3.6 and 4.1 $\mu$m (all thicknesses values are listed in caption of Fig. 2).

Figure 5 displays how the other ten $C_6H_6$ ice fundamental vibrational modes ($\nu_2$, $\nu_3$, $\nu_6$, $\nu_7$, $\nu_9$, $\nu_{10}$, $\nu_{11}$, $\nu_{16}$, $\nu_{17}$, and $\nu_{19}$) evolve with deposition temperature. These vibrations are much weaker in intensity than the fundamentals described above. The major result for this set of vibrational bands is the disappearance of several modes as the deposition temperature increases. This is the case for the $\nu_3$ and $\nu_{17}$ C-H symmetric bending modes, the $\nu_7$ and $\nu_{11}$ wagging vibrations, the $\nu_2$, and $\nu_{16}$ C–C stretching modes, whose bands peak at 1348 $cm^{-1}$, 1176 $cm^{-1}$, 991 $cm^{-1}$, 854 $cm^{-1}$, 999 $cm^{-1}$, and 1586 $cm^{-1}$ at 15 K, respectively. At temperatures higher than 15 K, these absorption bands decrease in intensity to disappear completely from 100 K. These specific vibrational modes are highly symmetric inactive modes in $C_6H_6$ gas. They are activated by the disorder of the asymmetric environment of the amorphous phase, and then turn into highly symmetric, again, in the ordered crystalline phase. Consequently, we identify this significant spectral change as an indication of the on-going crystallization of the $C_6H_6$ ice phase.

The four other fundamental vibrations $\nu_6$, $\nu_9$, $\nu_{10}$ and $\nu_{19}$ display different spectral changes with the deposition temperature. The $\nu_6$ C–C symmetric bending modes and the $\nu_9$ C–C stretching bands, which peak at 1011 $cm^{-1}$ and 1311 $cm^{-1}$ at 15 K, respectively, increase in intensity as the temperature increases and reach a maximum intensity between 120 K — 130 K. Between 15 K and 130 K, the vibration modes $\nu_6$ and $\nu_9$ shift to lower energies by 2 $cm^{-1}$ and to higher energies by 1 $cm^{-1}$, respectively. A small shoulder in the $\nu_9$ appears from 90 K to 130 K. The $\nu_{10}$ C–H in-plane bending mode peaks at 1146 $cm^{-1}$ at 15 K and splits into two maxima from 45 K. At 130 K, the asymmetric doublet has maxima at 1143 $cm^{-1}$ and 1148 $cm^{-1}$.

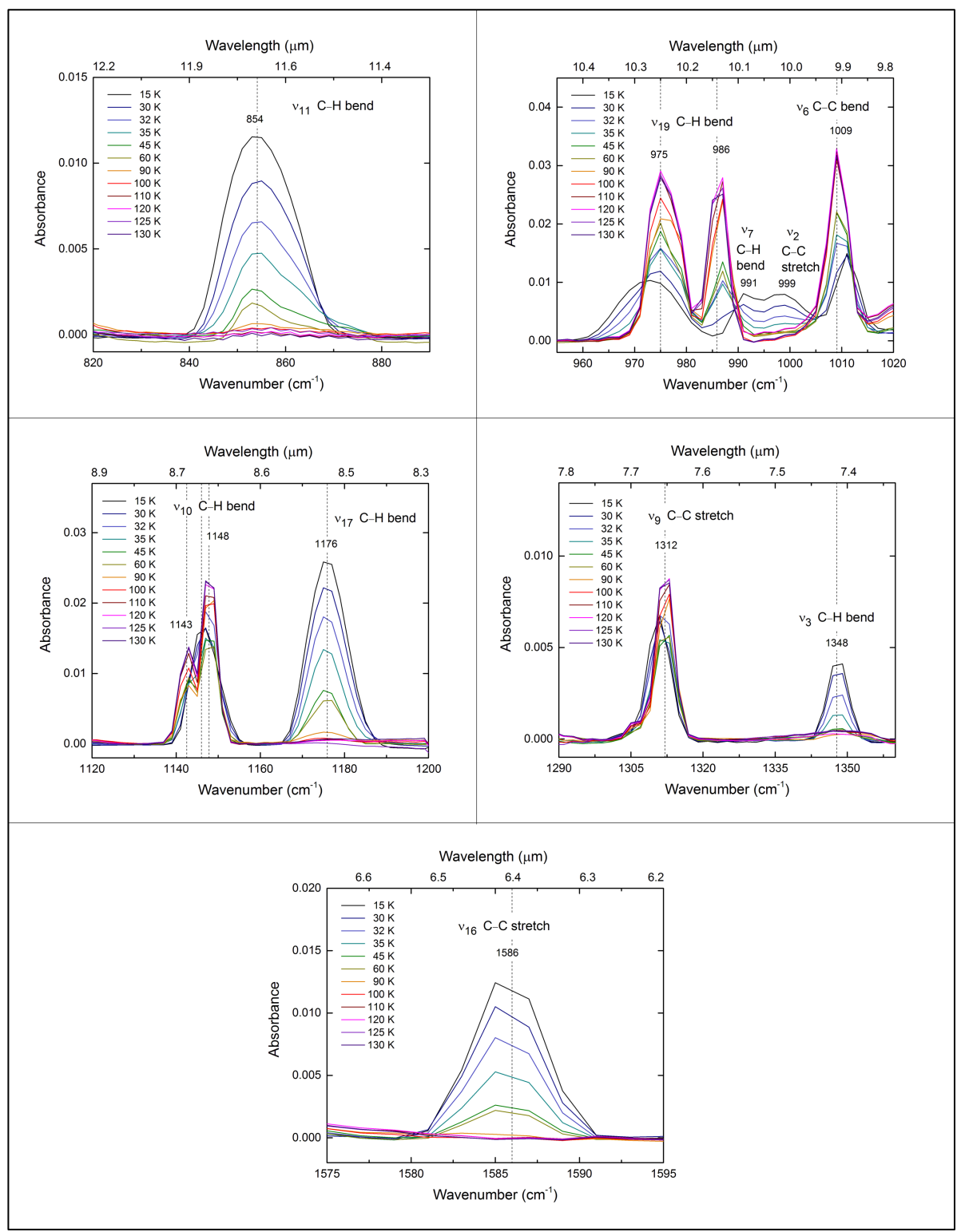

**Figure 5.** Other fundamental modes of $C_6H_6$ ice in the mid-IR region from absorbance spectra obtained after deposition of the benzene vapors at temperatures from 15 K to 130 K. Ice film thickness ranges between 3.6 and 4.1 $\mu m$ (all thicknesses values are listed in caption of Fig. 2).

The mid-IR spectra of $C_6H_6$ ice displays a large number of binary to quaternary combination bands (Table 1). Several of the combination bands, for which a large number of combinations is possible, have no spectral assignment, especially for the spectral region between 4550 $cm^{-1}$ and 6200 $cm^{-1}$. Figure 6 shows the four most intense combination bands and their temperature dependences. The remaining combination bands observed in the $C_6H_6$ ice spectra are very weak in intensity. As for the absorption bands presented in Figure 6, all the combination modes change markedly with the temperature. As the deposition temperature increases, several of them undergo splitting into two to four components with an increase in the band intensity and a frequency shift, as seen in Figure 6. While few of them experience a spectral shift to higher energies, like the three most intense combination bands at 1539 $cm^{-1}$, 1823 $cm^{-1}$ and 1966 $cm^{-1}$ at 15 K, the other combination modes displace to lower frequencies, as does the band at 4056 $cm^{-1}$ (Figure 6). At frequencies higher than 3620 $cm^{-1}$, all combination modes have a band shift towards lower energies.

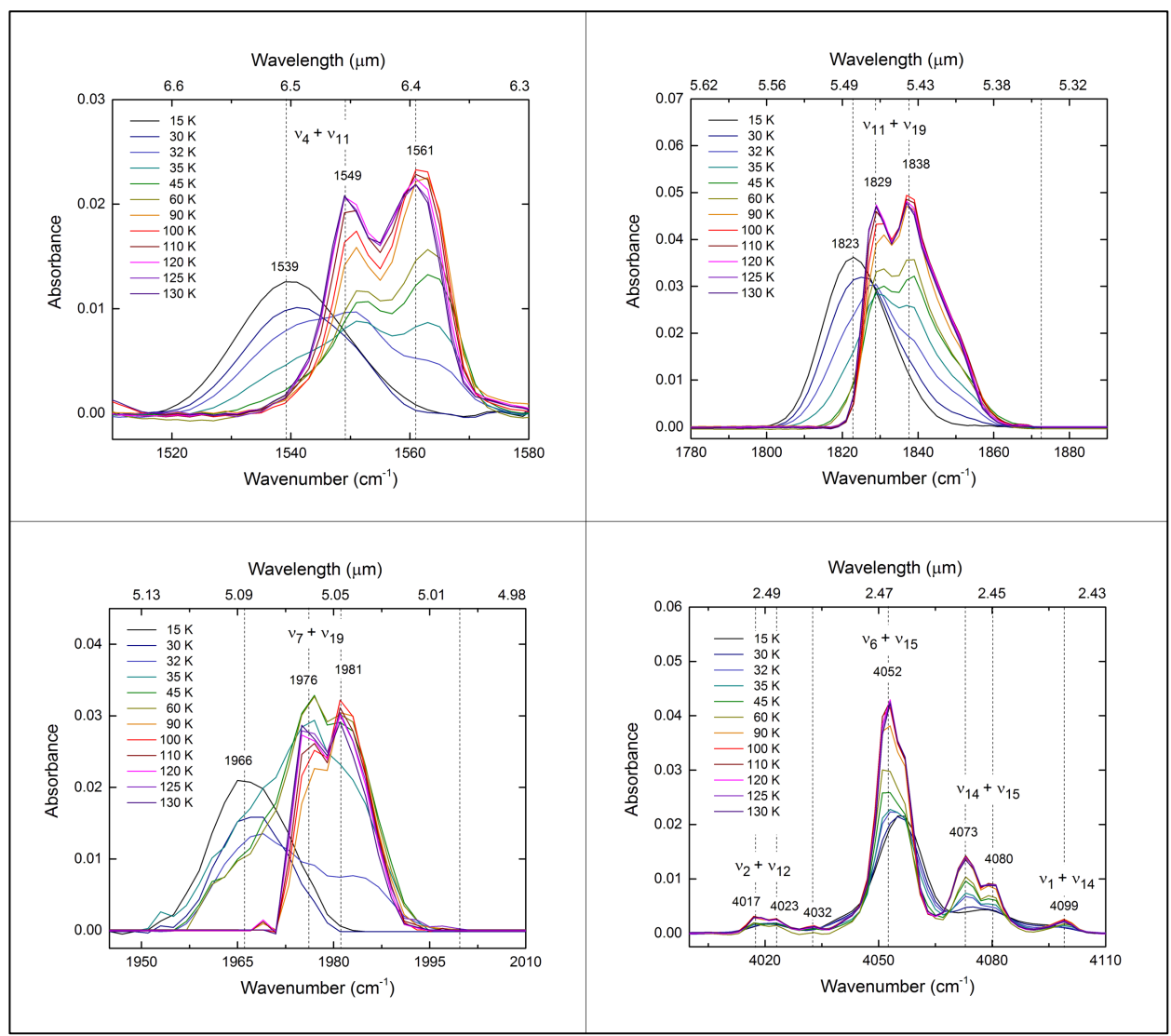

**Figure 6.** Most intense combination modes of $C_6H_6$ ice in the mid-IR region from absorbance spectra obtained after deposition of the benzene vapors at temperatures from 15 K to 130 K. Ice film thickness ranges between 3.6 and 4.1 $\mu$m (all thicknesses values are listed in caption of Fig. 2).

In the very high-frequencies region of the benzene ice spectra (frequencies higher than 5000 cm⁻¹), we observe the first overtone $2\nu_{CH}$ of the C–H stretching vibrations (Figure 7). This very weak overtone mode appears as a broad band at 15 K which peaks at 5980 cm⁻¹ and splits into several components as the temperature increases, with a main maximum at 5976 cm⁻¹ at 130 K.

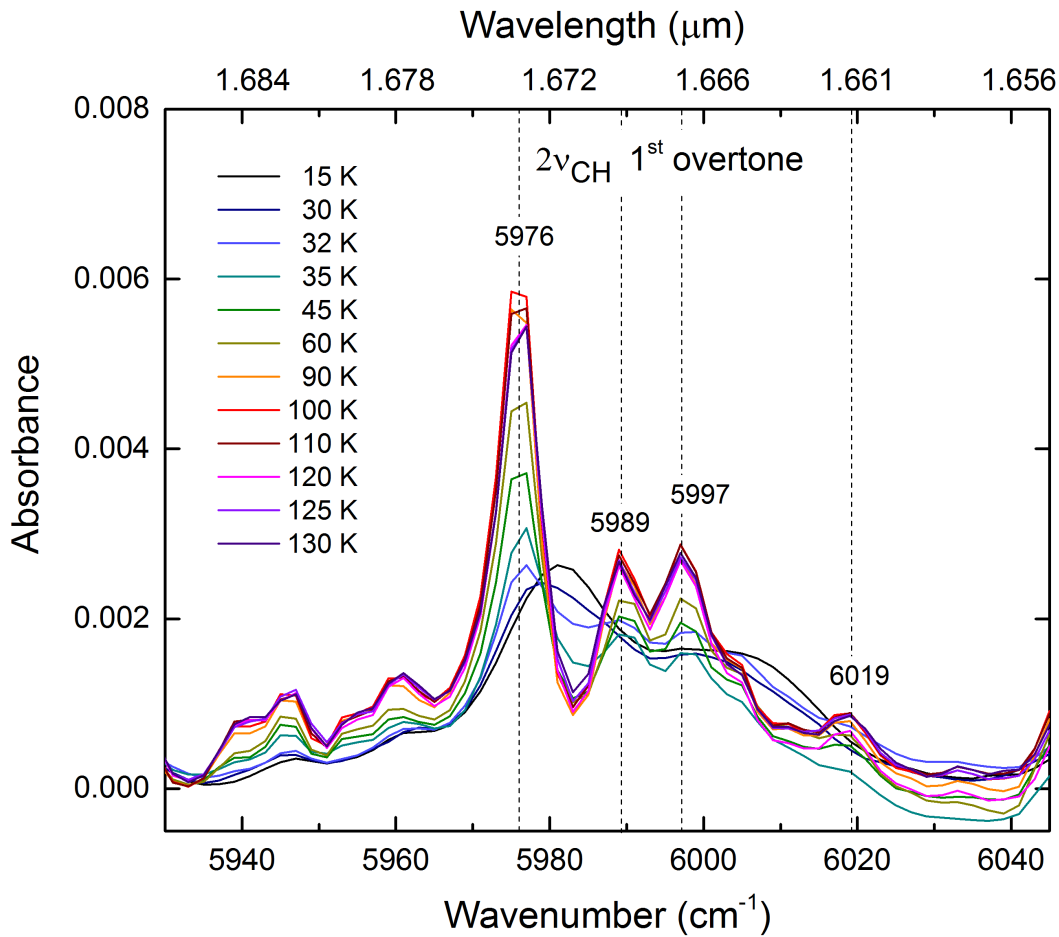

**Figure 7.** First overtone ($2\nu_{CH}$) of the C-H stretching vibrations of $C_6H_6$ icefrom absorbance spectra obtained after deposition of the benzene vapors at temperatures from 15 K to 130 K. Ice film thickness ranges between 3.6 and 4.1 μm (all thicknesses values are listed in caption of Fig. 2).

### 3.3. *Identification of $C_6H_6$ Ice Phases from the Experimental IR Spectra*

As described in the previous sections 3.1 and 3.2, the far- and mid-IR absorbance spectra of benzene ice reveal several temperature-dependent spectral changes. Except for the $\nu_{13}$ C-C stretching mode at 1478 cm$^{-1}$, all the $C_6H_6$ ice vibration modes are modified when $C_6H_6$ vapor is deposited at different temperatures. The increase of the deposition temperature from 15 K to 130 K triggers band sharpening and/or band splitting, an increase in band intensity, and shifts in band position. As well, as noted in section 3.2, some $C_6H_6$ ice absorption bands at 15 K disappear at temperatures ≥ 100 K and thus behave as infrared-inactive in the crystal. This temperature-dependence behavior evidences the transition of $C_6H_6$ ice from the amorphous to crystalline phase, as generally observed for ices studied in laboratory. In our experiments, we identify the amorphous phase of $C_6H_6$ ice at 15 K and 30 K for which the IR spectra presents only singlet vibrational bands that do not undergo any frequency shift. From 32 K, we notice that various spectral changes are initiated. In the far-IR spectral region, the behavior of the $\nu_{20}$ ring deformation mode of $C_6H_6$ ice, which changes from a broad singlet at 15 K and 30 K to a doublet at 32 K, is a strong indication that the molecules of $C_6H_6$ ice start to order. In the mid-IR region, several fundamental vibration modes and combination modes start to split at 32 K (appearance of a shoulder or inflection in the bands). From 32 K to 120 K, the far- and mid-IR vibrational bands of benzene ice evolve significantly with the temperature and produce distinct transitions in the spectra that illustrate the on-going reordering of the molecules towards a lower energy state. From 120 K to 130 K, $C_6H_6$ ice absorption bands do not undergo any further spectral changes, resulting in identical IR spectra and indicating that the ordering and crystal orientations of $C_6H_6$ ice molecules are complete. At temperatures higher than 130 K, the ice samples sublime, which alters the mid- and far-IR spectra with a decrease in the intensity of all $C_6H_6$ absorption bands.

A previous laboratory study on benzene ice, directly deposited at low temperatures from 13 K to 78 K or annealed up to 96 K, reports a transition in the ice phase from amorphous to crystalline from around 60 K using Raman spectroscopy and X-ray diffraction (Ishii et al. 1996). Ishii et al. report no significant change in the Raman spectra and diffraction patterns for temperatures lower than 60 K, and they define the $C_6H_6$ ice crystallization temperature around 60 K. Mouzay et al. (2021), after depositing at 16 K and warming it up to 300K, observe the evolution of the surface coverage of the $\nu_{13}$ and $\nu_{14}$ vibrational modes of $C_6H_6$ ice with temperature, and report similarly benzene crystallization at around 55 K. Hollenberg & Dows (1962) characterize $C_6H_6$ ice at 85 K as crystalline. On the other hand, differential thermal analyses (DTA) of benzene have determined the glass transition temperature of benzene to be at 120 K (Dubochet et al. 1984) or at 130 K (Angell et al. 1978). And results from nuclear magnetic resonance (NMR) absorption spectroscopy of solid $C_6H_6$ from 75 K to 278 K have identified that the reorientation of $C_6H_6$ molecules about their six-axes is initiated at about 90 K (Andrew & Eades, 1953). In this present study, we assessed the crystalline temperature (i.e. the temperature at which the crystallization is completely achieved) based on two criteria: 1) there are no observed changes (or almost no changes) in the strength and/or spectral dependence of the IR absorption bands after an increase in the deposition temperature, and 2) there is no noticeable sublimation of the ice film at the temperature held after vapor deposition. Based on these criteria and on our results of the temperature dependence of the far and mid-IR spectra of benzene ice, we have identified that $C_6H_6$ ice has completed crystallization at the deposition temperature ($T_d$) determined as $120$ K $\leq T_d \leq 130$ K. This result is compatible with the glass transition temperatures previously estimated. For this confined temperature range, the crystalline phase of benzene ice is unambiguously identified from the disordered amorphous and ordering transition phases.

### 3.4. *Optical Constants of Amorphous and Crystalline $C_6H_6$ Ices*

Figure 8 displays the real ($n$) and imaginary ($k$) parts of the ice complex refractive index of benzene ice in the far and mid-IR from 50 cm$^{-1}$ to 8000 cm$^{-1}$ ($200 - 1.25$ µm). We calculated $n$ and $k$ from our laboratory spectra at each temperature from 15 K to 130 K, and using $n_0$ value of $1.54 \pm 0.02$ determined by Romanescu et al. (2010). The complex indices of refraction of $C_6H_6$ ice can be used to determine absorption cross-sectional spectra, which can be compared with astronomical emission spectra. We are the first to report cryogenic measurements of $n$ and $k$ of $C_6H_6$ ice at a variety of temperatures in both the amorphous and crystalline phases. We observe that $n$ and $k$ vary with the temperature (Figures 10 and 11).

In Figure 9, we illustrate the spectral dependence of $n$ and $k$ for the most intense vibrational mode of $C_6H_6$ ice, the $\nu_4$ C–H asymmetric bending mode, in the amorphous phase at 15 K (black curve) and in the crystalline phase at 130 K (red curve).

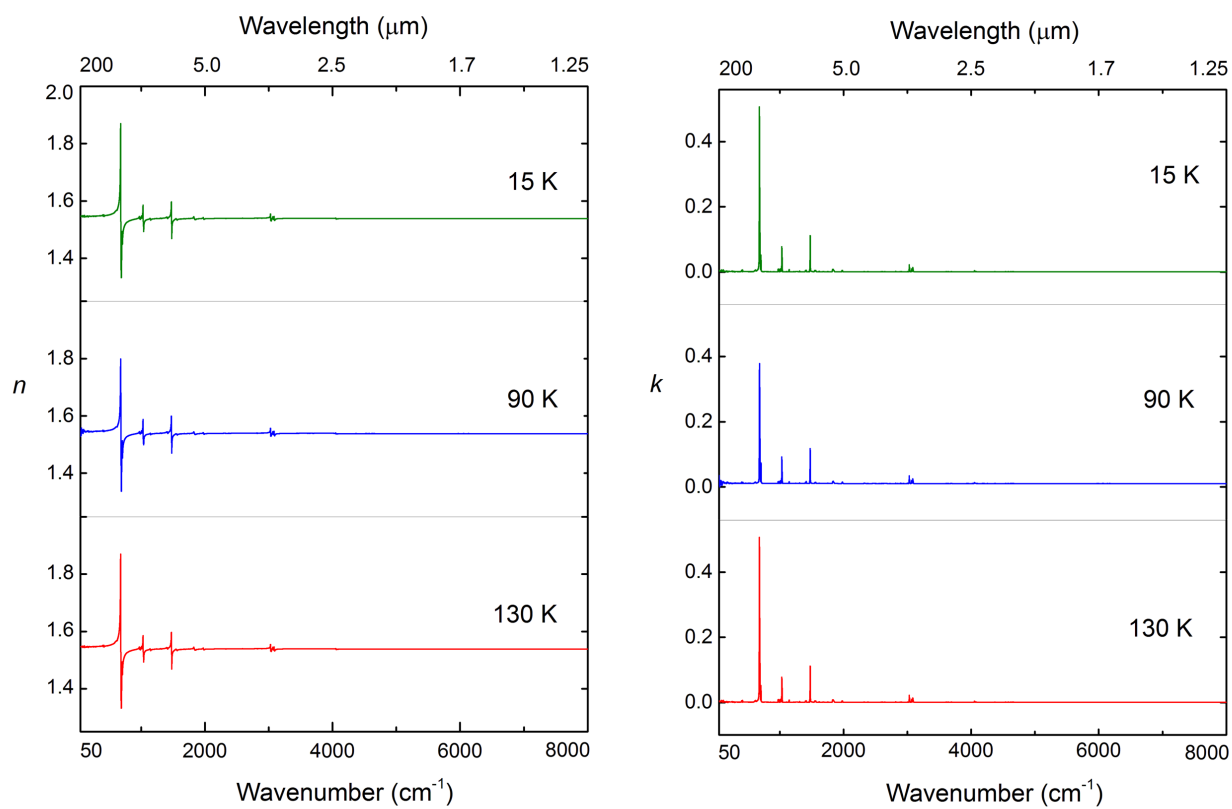

**Figure 8.** Spectral dependence of the real ($n$) and imaginary ($k$) parts of the complex refractive index of $C_6H_6$ ice (4 $\mu$m thick film) at deposition temperatures of 15 K, 90 K and 130 K in the spectral range 50 − 8000 cm$^{-1}$ (200 − 1.25 $\mu$m). $C_6H_6$ ice was formed by direct vapor deposition at 15 K, 90 K and 130 K to produce thin ice films of 3.87 $\mu$m, 3.71 $\mu$m, and 3.63 $\mu$m, respectively. $n$ and $k$ were calculated using a custom-developed IDL program which uses the Beer-Lambert law and an iterative Kramers-Kronig analysis of the laboratory corrected absorbance spectra of $C_6H_6$ ice.

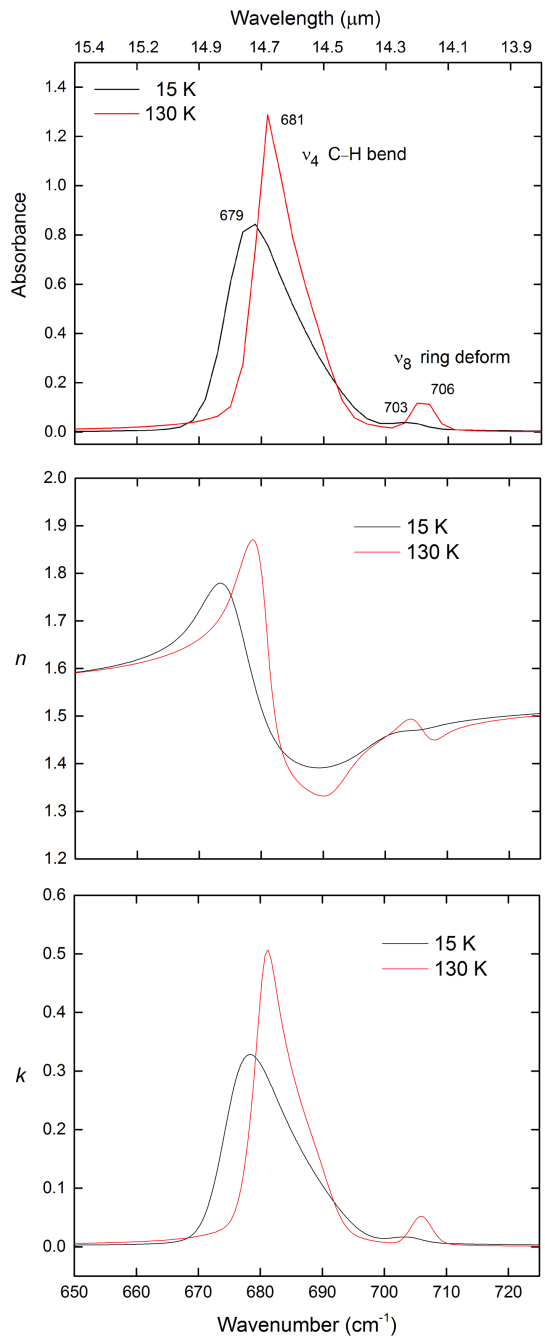

**Figure 9.** Absorbance spectra and associated optical constants (real part *n* and imaginary part *k* of the complex refractive index) for the $\nu_4$ C–H asymmetric bending mode of $C_6H_6$ ice in the amorphous phase at 15 K (*black curve*) and in the crystalline phase at 130 K (*red curve*). $C_6H_6$ ice was formed by direct vapor deposition at 15 K and 130 K to produce thin ice films of 3.87 $\mu$m and 3.63 $\mu$m, respectively, that were analyzed by FTIR spectroscopy from 50 cm$^{-1}$ to 8000 cm$^{-1}$ (200 − 1.25 $\mu$m).

## 4. IMPLICATIONS TO TITAN STUDIES

As indicated in Section 1, Vinatier et al. (2018) have recently identified benzene ice mixed with other unidentified compounds in Titan's stratosphere by analyzing Cassini CIRS nadir and limb mid-IR spectra. From limb observations acquired in March 2015, they detected an ice cloud emission feature peaking at 682 cm$^{-1}$ at altitudes between 168 and 278 km, and identified solid benzene as the best chemical match of this emission feature. For their study, they used the experimental spectrum of crystalline $C_6H_6$ ice deposited at 130 K by Schmitt et al. (2015). Specifically, they analyzed the $C_6H_6$ $\nu_4$ C–H out-of-plane bending mode detected in the IR spectrum of Schmitt et al. (2015), to fit the 682 cm$^{-1}$ emission feature and they determined the spectral dependences of the extinction, absorption, and scattering cross sections per unit particle volume of pure $C_6H_6$ ice. In the Schmitt et al. (2015) experiment, a 1.65 µm thin film of $C_6H_6$ ice, produced by vapor deposition of $C_6H_6$ vapors at 130 K, was analyzed in the mid-IR region from 400 cm to 4200 cm$^{-1}$ (25 — 2.38 µm). In their mid-IR spectra, the $\nu_4$ C–H out-of-plane bending mode appeared as an asymmetric doublet band with the two maxima at 679 cm$^{-1}$ and 681 cm$^{-1}$ and a width of ~6.5 cm$^{-1}$. Vinatier et al. (2018) indicated that in the Schmitt et al. (2015) experimental $C_6H_6$ spectra, the most intense bands of $C_6H_6$ ice were very weakly sensitive with temperature, with an observed spectral shift < 0.5 cm$^{-1}$ between 60 K and 130 K. We, however, were unable to corroborate this result since we did not find any published work from Schmitt et al. (2015) reporting a study on the spectral and temperature dependence of $C_6H_6$ ice. From personal communication, we have been acquainted that Schmitt and colleagues carried out this experimental study by condensing $C_6H_6$ at 130K and then cooling to temperatures down to 60 K. Our results displayed in Figure 11 differ slightly from data obtained by Schmitt et al. (2015). In our mid-IR spectra, the $\nu_4$ vibrational mode at 130 K appears as a singlet band that is less intense than the doublet observed by Schmitt et al. (2015). In our studies, this band peaks at 681 cm$^{-1}$ and has a width of ~7.5 cm$^{-1}$ (Figure 10). Mouzay et al. (2021) reported in their $C_6H_6$ crystalline ice spectra at 130 K as well a single band for the $\nu_4$ C–H out-of-plane bending mode, but peaking at 679 cm$^{-1}$. From condensation temperatures from 60 K to 130 K, we observe a band displacement of 2 cm$^{-1}$ to higher energies and also that the $\nu_4$ vibrational mode is temperature-dependent (Figure 4). As pointed out in section 3.2, we observe that as the condensation temperature increases, the $\nu_4$ band becomes sharper with an increase in band intensity and shifts to higher energies.

Mouzay et al. (2021) showed that the $\nu_4$ vibrational mode is shifted to higher frequencies from 16 K to 130 K by 5 cm$^{-1}$ but, contrarily to our results, its shape looks almost identical. And after depositing the vapor of $C_6H_6$ on a gold-plated substrate at 130 K, keeping at 130 K for a few hours and finally cooling down to 70 K, no significant frequency shift in $\nu_4$ band is observed with temperature and the band intensity very slightly varies from 70 K to 130 K.

The discrepancies in temperature dependences between our spectral results, Schmitt et al. unpublished work, and Mouzay et al. (2021) certainly result in the fact that we have studied IR spectra of $C_6H_6$ condensed (or deposited) at different temperatures, whereas Schmitt et al. (2015) and Mouzay et al. (2021) compared the spectra of a sample condensed at 130 K and then cooled down to 60 K, or 70 K, respectively. While we have compared how the temperature affect the state of crystallization as a function of deposition temperature, Schmitt et al. (2015) and Mouzay et al. (2021) studied the reversible temperature effects upon cooling-warming their crystallized $C_6H_6$ ice at temperature below its condensation temperature.

The difference in the shape of the band between our experimental study and the work of Schmitt et al. (2015) in which $C_6H_6$ is condensed at 130 K is certainly the result of the distinct spectral resolution of the data used in both experiments, 4 cm$^{-1}$ in our work, while 1 cm$^{-1}$ in Schmitt et al. (2015). The bands may not fully be resolved at 4 cm$^{-1}$ resolution and can induce spectral shape changes such as wider and weaker bands and even some spectral shifts of the order of up to 1 cm$^{-1}$ if the band is asymmetric or if there exists a weaker side band or a shoulder at a separation less than the spectral resolution. This can explain that our experimental spectra display only a single band at 681 cm$^{-1}$, and maybe only part of the shift. As well, the width and intensity difference of the narrow 706 cm$^{-1}$ band in Schmitt et al. (2015) and our wider less intense $\nu_8$ ring deformation mode. However, it is difficult to fully explain the shift of the $\nu_4$ C-H out-of-plane bending mode at 681 cm$^{-1}$ and its strong difference in intensity between our work and Schmitt et al. (2015). Both states of crystallization may not be exactly the same. In our experiments, we used diamond as the substrate material while than Schmitt et al. (2015) used cesium iodide (CsI) which may also play a role in the crystallization. We note that, except using a different substrate material, we have no explanation for the spectral differences observed between Mouzay et al. (2021) results (single band of $\nu_4$ at 679 cm$^{-1}$) and Schmitt et al. (2015) ($\nu_4$ asymmetric doublet band at 679 and 681 cm$^{-1}$), even if using a similar laboratory technique and the same spectral resolution of 1 cm⁻¹ to obtain crystalline $C_6H_6$ ices and look at their temperature spectral dependence.

Compared to the Vinatier et al. (2018) $C_6H_6$ ice cloud emission feature at 682 cm⁻¹, our data at 130 K gives a peak frequency of the $\nu_4$ band of crystalline $C_6H_6$ ice 1 cm⁻¹ lower in energy.

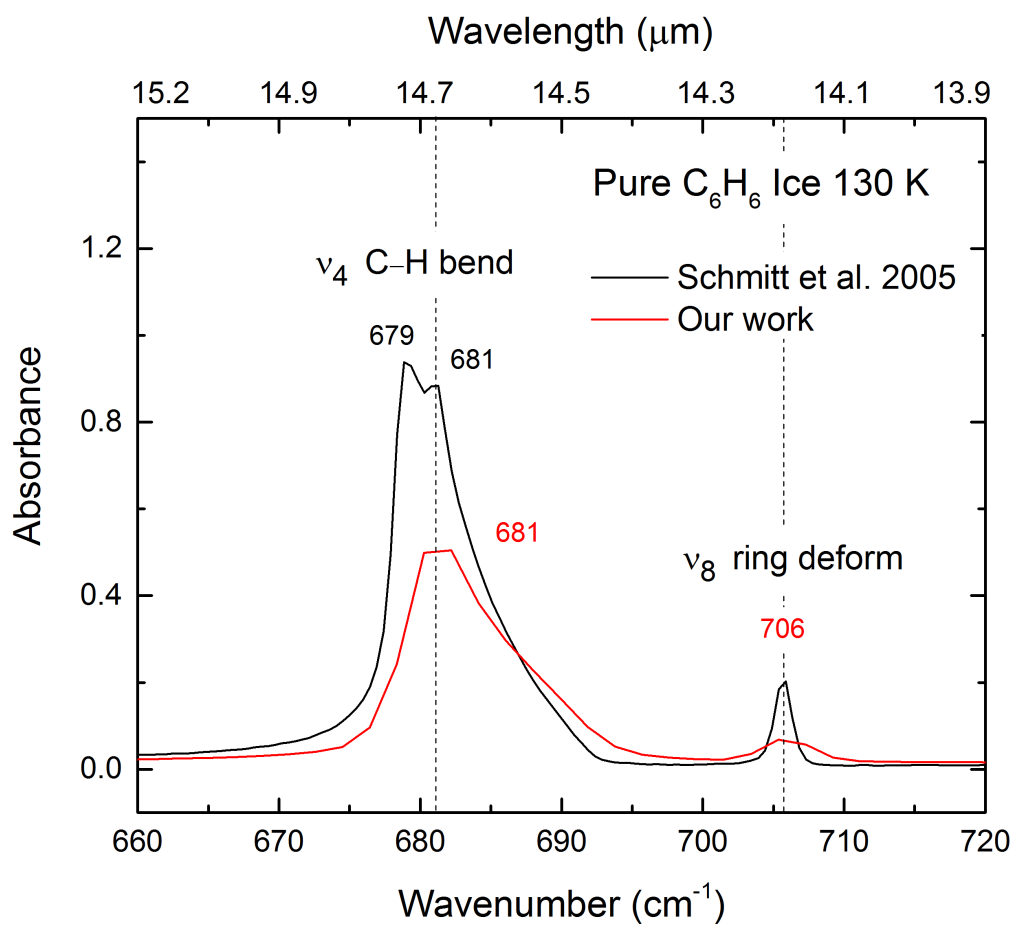

**Figure 10.** Comparisons of laboratory absorbance spectra of crystalline $C_6H_6$ ice from Schmitt et al. (2015) and our data obtained under similar experimental conditions using the SPECTRAL chamber (current work). The black curve shows the Schmitt et al. (2015) absorbance spectrum at 130 K obtained from a $C_6H_6$ thin ice film of 1.65 μm using a spectral resolution of 1 cm⁻¹, while the red curve depicts the absorbance spectrum from this work recorded at a spectral resolution of 4 cm⁻¹, in which $C_6H_6$ vapor was deposited at 130 K to form a thin ice film of 1.65 μm. Superimposed in each figure are the vibrational transitions and peak frequencies for each observed ice absorption band.

Schmitt et al. (2015) have provided as well the real and imaginary parts of the ice complex refractive index of benzene ice at 130 K from 400 to 4200 $cm^{-1}$ available in the SSHADE[4] database. Complex refractive indices, which are temperature-dependent (as we show in Figures 9 and 10), are crucial to determine the abundances of condensed species observed in planetary atmospheres. Moreover, for low-temperature environments, such as Titan's atmosphere and surface, where thermal energy is at a minimum, it is essential to obtain experimental optical data in the far-IR spectral region that has high thermal emission below 400 $cm^{-1}$, where CIRS has measured in detail during the 13-year Cassini mission, and for which optical constants of organic ices observed in Titan's stratosphere are still lacking. In our work, we have generated a wide database of optical constants of benzene ice from 50 $cm^{-1}$ up to 8000 $cm^{-1}$ in order to provide critically-missing experimental spectroscopic and optical data for $C_6H_6$ ice needed to continue interpreting Cassini-observed spectra of Titan's atmosphere.

Regarding the CIRS-observed HASP ice cloud detected by Anderson et al. (2018a) in Titan's late southern fall stratosphere at high southern latitudes, its emission feature peaks near 200 $cm^{-1}$ (see Figure 19 in Anderson et al. 2018a). The vertical extent of the HASP ice cloud is observed at stratospheric altitudes where pure $C_6H_6$, pure hydrogen cyanide (HCN), and pure cyanoacetylene ($HC_3N$) are expected to condense and form stratospheric ice clouds (see Figure 1). In Titan's atmosphere, most of the organic vapors will condense to form ice shells on aerosol solid particles as the vapors cool while descending throughout Titan's stratosphere. The $C_6H_6$ ice spectra that we have obtained in this study did not display any detectable absorption bands between 50 — 400 $cm^{-1}$ (100 - 25 µm), for any of the deposition temperatures studied (15 — 130 K). With our instrument and experimental protocol, we are able to detect absorption bands at this wavenumber range with absorbance height as low as 0.002 absorbance unit (for S/N>3). For example, we detected the weak far-IR absorption bands of propionitrile between 100 $cm^{-1}$ and 390 $cm^{-1}$ (see papers of Nna-Mvondo et al. 2019, Anderson et al., 2018a). This was not the case for $C_6H_6$ ice. The HASP ice cloud has a very intense and broad far-IR spectral feature centered at 200 $cm^{-1}$ (Anderson et al., 2018a). Compared with those previous measurement of other molecules absorbing in this frequency range integrated with other observed ice cloud bands that we carried out with the same experimental setup and method (Anderson et al. 2018a,b, Nna-Mvondo et al. 2019), we should be able to detect such intense and wide band around 200 cm$^{-1}$ if it was present for $C_6H_6$ ice. Besides, other published far-IR spectral study of solid benzene (Chantry et al. 1967, Harada & Shimanouchi 1967, 1971, Sataty et al. 1973, Sataty & Ron 1976) did not detect neither any absorption band around 200 cm$^{-1}$. So we agree that we can rule out $C_6H_6$ ice as the only absorber of the HASP ice cloud spectral feature.

Therefore, our results definitely rule out pure $C_6H_6$ ice as the spectral match of the HASP cloud emission feature. However, as reported in Anderson et al. (2018a), the HASP ice cloud could be formed by a co-condensation scenario, in which Titan's organic vapors, including benzene, enter altitude regions in the stratosphere where the HASP cloud is observed and where they can simultaneously saturate to form a mixed (or co-condensed) ice cloud. In addition to the HASP ice cloud, ice cloud formation processes in Titan's stratosphere via vapor co-condensation have already been reported for the formation of the northern winter 160 cm$^{-1}$ ice cloud observed in Titan's northern winter lower stratosphere (Anderson & Samuelson 2011; Anderson et al. 2018b), which was experimentally determined to contain (at a minimum) co-condensed HCN and $HC_3N$.

## 5. CONCLUSION

To the best of our knowledge, our work is the first publication of far- and mid-IR absorbance spectra of benzene ice conducted over a large range of deposition temperatures. We have provided the associated optical constants for each of the studied temperatures from 15 K to 130 K. These spectra and optical constants generated from cold to warmer temperatures are valuable data for research studies related to Titan's stratosphere, but as well to Titan's surface. Indeed, solid benzene has been tentatively identified on Titan's surface (Niemann et al. 2005; Clark et al. 2010). Our low temperature data of $C_6H_6$ ices (< 80 K) are also useful for the investigation of other cold astronomical environments where benzene has been observed, such as in the atmospheres of Jupiter (Kim et al. 1985), Saturn (Bézard et al. 2001), and in the proto-planetary nebula CRL 618 (Cernicharo et al. 2001).

ACKNOWLEDGMENTS

D.N.-M and C.M.A. acknowledge research funding support by the NASA Internal Scientist Funding Model (ISFM) through the Fundamental Laboratory Research (FLaRe) work package.

REFERENCES

Allamandola, L. J., Tielens, A. G. G. M., & Barker, J. R. 1989, The Astrophysical Journal Supplement Series, 71, 733

Anderson, C. M., & Samuelson, R. E. 2011, Icarus, 212, 762

Anderson, C. M., Samuelson, R. E., Achterberg, R. K., Barnes, J. W., & Flasar, F. M. 2014, Icarus, 243, 129

Anderson, C. M., Samuelson, R. E., & Nna-Mvondo, D. 2018a, Space Science Reviews, 214, 125

Anderson, C. M., Nna-Mvondo, D., Samuelson, R. E., McLain, J. L., & Dworkin, J. P. 2018b, The Astrophysical Journal, 865, 62

Andrew, E. R., & Eades, R. G. 1953, Proceedings of the Royal Society of London Series A Mathematical and Physical Sciences, 218, 537

Angell, C. A., Sare, J. M., & Sare, E. J. 1978, The Journal of Physical Chemistry, 82, 2622

Bertie, J. E., & Keefe, C. D. 2004, Journal of Molecular Structure, 695-696, 39

Bézard, B., Drossart, P., Encrenaz, T., & Feuchtgruber, H. 2001, Icarus, 154, 492

Bittner, J. D., & Howard, J. B. 1981, Symposium (International) on Combustion, 18, 1105

Bonadeo, H., Marzocchi, M. P., Castellucci, E., & Califano, S. 1972, The Journal of Chemical Physics, 57, 4299

Buss, R. H., Jr., Tielens, A. G. G. M., Cohen, M., Werner, M. W., Bregman, J. D., & Witteborn, F. C. 1993, The Astrophysical Journal, 415, 250

Callahan, M. P., Gerakines, P. A., Martin, M. G., Peeters, Z., & Hudson, R. L. 2013, Icarus, 226 1201

Cernicharo, J., Heras, A. M., Tielens, A. G. G. M., Pardo, J. R., Herpin, F., Guélin, M., & Waters, L. B. F. M. 2001, The Astrophysical Journal, 546, L123

Chantry, G. W., Gebbie, H. A., Lassier, B., & Wyllie, G. 1967, Nature, 214, 163

Cherchneff, I. 2011, EAS Publications Series, 46, 177

Clark, R. N., et al. 2010, Journal of Geophysical Research: Planets, 115

Clemett, S. J., et al. 1994, Meteoritics, 29, 457

Coates, A. J., Crary, F. J., Lewis, G. R., Young, D. T., Waite Jr., J. H., &
    Sittler Jr., E. C. 2007, Geophysical Research Letters, 34

Coustenis, A., et al. 2013, The Astrophysical Journal, 779, 177

Coustenis, A., et al. 2016, Icarus, 270, 409

Coustenis, A., et al. 2018, The Astrophysical Journal, 854, L30

Domingo, M., Millán, C., A, S. M., & Canto, J. 2007, Proceedings SPIE 6616,
    Optical Measurement Systems for Industrial Inspection V, 66164A

Dubochet, J., Alba, C. M., MacFarlane, D. R., Angell, C. A., Kadiyala, R. K.,
    Adrian, M., & Teixeira, J. 1984, The Journal of Physical Chemistry, 88, 6727

Epstein, H., & Steiner, W. 1934, Nature, 133, 910

Frenklach, M., & Feigelson, E. D. 1989, The Astrophysical Journal, 341, 372

Hahn, J. H., Zenobi, R., Bada, J. L., & Zare, R. N. 1988, Science, 239, 1523

Harada, I., & Shimanouchi, T. 1967, The Journal of Chemical Physics, 46, 2708,
    The Journal of Chemical Physics, 55, 3605

Hayatsu, R., Matsuoka, S., Scott, R. G., Studier, M. H., & Anders, E. 1977,
    Geochimica et Cosmochimica Acta, 41, 1325

Herzberg, G. 1945, Molecular spectra and molecular structure. Vol.2: Infrared
    and Raman spectra of polyatomic molecules

Hirschfeld, T., & Mantz, A. W. 1976, Appl Spectrosc, 30, 552

Hollenberg, J. L., & Dows, D. A. 1962, The Journal of Chemical Physics, 37,
    1300

Ishii, K., Nakayama, H., Yoshida, T., Usui, H., & Koyama, K. 1996, Bulletin of
    the Chemical Society of Japan, 69, 2831

Khanna, R. K., Ospina, M. J., & Zhao, G. 1988, Icarus, 73, 527

Khare, B. N., Bakes, E. L. O., Imanaka, H., McKay, C. P., Cruikshank, D. P., &
    Arakawa, E. T. 2002, Icarus, 160, 172

Kim, S. J., Caldwell, J., Rivolo, A. R., Wagener, R., & Orton, G. S. 1985,
    Icarus, 64, 233

Kim, Y. S., & Kaiser, R. I. 2009, The Astrophysical Journal Supplement Series,
    181, 543

Lebonnois, S., Bakes, E. L. O., & McKay, C. P. 2002, Icarus, 159, 505

Leger, A., & Puget, J. L. 1984, Astronomy and Astrophysics, 500, 279

Mair, R. D., & Hornig, D. F. 1949, The Journal of Chemical Physics, 17, 1236

Materese, C. K., Nuevo, M., & Sandford, S. A. 2015, The Astrophysical Journal,
    800, 8pp

Miani, A., Cané, E., Palmieri, P., Trombetti, A., & Handy, N. C. 2000, The
    Journal of Chemical Physics, 112, 248

Mimura, K. 1995, Geochimica et Cosmochimica Acta, 59, 579

Mouzay, J., Couturier-Tamburelli, I., Piétri, N., & Chiavassa, T. 2021, Journal of Geophysical Research: Planets, 126, e2020JE006566

Niemann, H. B., et al. 2005, Nature, 438, 779

Nna-Mvondo, D., Anderson, C. M., & Samuelson, R. E. 2019, Icarus, 333, 183

Oliver, D. A., & Walmsley, S. H. 1969, Molecular Physics, 17, 617

Pering, K. L., & Ponnamperuma, C. 1971, Science, 173, 237

Rocha, W., & Pilling, S. 2014, Spectrochimica Acta Part A: Molecular and Biomolecular Spectroscopy, 123, 436

Romanescu, C., Marschall, J., Kim, D., Khatiwada, A. and Kalogerakis, K.S. 2010, Icarus, 205(2), 695

Sagan, C., et al. 1993, The Astrophysical Journal, 414, 399

Sataty, Y. A., & Ron, A. 1976, The Journal of Chemical Physics, 65, 1578

Sataty, Y. A., Ron, A., & Brith, M. 1973, Chemical Physics Letters, 23, 500

Schinder, P. J., Flasar, F. M., Marouf, E. A., French, R. G., McGhee, C. A., Kliore, A. J., Rappaport, N. J., Barbinis, E., Fleischman, D., & Anabtawi, A., 2011, Icarus, 215, 460

Schmitt, B., Vinatier, S., Bernard, J.M. 2015, Mid-IR transmission and optical constants spectra of crystalline C6H6 at 130K. SSHADE/GhoSST (OSUG Data Center).                        Dataset/Spectral                        Data. https://doi.org/10.26302/SSHADE/EXPERIMENT_BS_20170830_001

Shock, E. L., & Schulte, M. D. 1990, Nature, 343, 728

Simoneit, B. R. T., & Fetzer, J. C. 1996, Organic Geochemistry, 24, 1065

Strazzulla, G., & Baratta, G. A. 1991, Astronomy and Astrophysics, 241, 310

Teanby, N. A., Sylvestre, M., Sharkey, J., Nixon, C. A., Vinatier, S., & Irwin, P. G. J. 2019, Geophysical Research Letters, 46, 3079

Tempelmeyer, K. E., & Mills, D. W., Jr. 1968, Journal of Applied Physics, 39, 2968

Tielens, A. G. G. M., & Charnley, S. B. 1997, in Planetary and Interstellar Processes Relevant to the Origins of Life, ed. D. C. B. Whittet (Dordrecht: Springer Netherlands), 23

Tielens, A. G. G. M. 2013, in Planets, Stars and Stellar Systems: Volume 5: Galactic Structure and Stellar Populations, eds. T. D. Oswalt, & G. Gilmore (Dordrecht: Springer Netherlands), 499

Trainer, M. G., Pavlov, A. A., Jimenez, J. L., McKay, C. P., Worsnop, D. R., Toon, O. B., & Tolbert, M. A. 2004, Geophysical Research Letters, 31

Vinatier, S., et al. 2018, Icarus, 310, 89

Vuitton, V., Yelle, R. V., & Cui, J. 2008, Journal of Geophysical Research: Planets, 113

Waite, J. H., Young, D. T., Cravens, T. E., Coates, A. J., Crary, F. J., Magee,
    B., & Westlake, J. 2007, Science, 316, 870

Wang, H., & Frenklach, M. 1997, Combustion and Flame, 110, 173

Wilson Jr., E.B. 1934, Physical Review, 45, 706

Wilson, E.B., Decius, J.C., Cross, P.C. 1955, Molecular vibrations: the theory
    of infrared and Raman vibrational spectra, McGraw-Hill, New York

Wilson, E. H., Atreya, S. K., & Coustenis, A. 2003, Journal of Geophysical
    Research: Planets, 108

Wyatt, R. E. 1998, The Journal of Chemical Physics, 109, 10732

Zhou, L., Kaiser, R. I., & Tokunaga, A. T. 2009, Planetary and Space Science,
    57, 830